\documentclass[lettersize,journal]{IEEEtran}

\usepackage{xcolor,soul,framed} %,caption

\colorlet{shadecolor}{yellow}
\usepackage{graphicx}
\graphicspath{{../pdf/}{../jpeg/}}
\DeclareGraphicsExtensions{.pdf,.jpeg,.png}

\usepackage[cmex10]{amsmath}
\usepackage{array}
\usepackage{mdwmath}
\usepackage{mdwtab}
\usepackage{eqparbox}
\usepackage{url}
\usepackage{cite}
\usepackage{stfloats}
\usepackage{subfigure}
\usepackage{float}
\usepackage{ragged2e}
\usepackage{caption}
\usepackage{setspace}
\usepackage{color}
\usepackage{comment}
\usepackage{booktabs}
\usepackage[utf8]{inputenc}
\usepackage[T1]{fontenc}
\usepackage{indentfirst}
\usepackage{makecell}
\usepackage[export]{adjustbox}
\usepackage{wrapfig}
\usepackage{algorithm}  
\usepackage{algpseudocode}  
\usepackage{amsmath}

\usepackage{multirow}
\usepackage{amssymb}

\usepackage{tikz,xcolor}
\usepackage[implicit=false]{hyperref}

\hypersetup{hidelinks,
	colorlinks=true,
	allcolors=black,
	pdfstartview=Fit,
	breaklinks=true}
\definecolor{lime}{HTML}{A6CE39}
\DeclareRobustCommand{\orcidicon}{
\begin{tikzpicture}
\draw[lime, fill=lime] (0,0)
circle[radius=0.16]
node[white]{{\fontfamily{qag}\selectfont \tiny \.{I}D}};
\end{tikzpicture}
\hspace{-2mm}
}
\foreach \x in {A, ..., Z}{%
\expandafter\xdef\csname orcid\x\endcsname{\noexpand\href{https://orcid.org/\csname orcidauthor\x\endcsname}{\noexpand\orcidicon}}
}

\newcommand{\circled}[1]{{\normalsize{\textcircled{\scriptsize{#1}}}\normalsize\;}}

\newcommand{\hhw}[1]{{\color{black}#1}} %COMMENTS BY Huawei
\newcommand{\hw}[1]{{\color{black}#1}} %COMMENTS BY Huawei for 1st-round review, 2024-Oct.
\newcommand{\secondReply}[1]{{\color{black}#1}} %COMMENTS BY Ye Guang for 2nd-round review, 2024-Dec-23.

\newcommand{\Guang}[1]{{\color{black}#1}}  %COMMENTS BY Ye Guang

\begin{document}

\title{BlockEmulator: An Emulator Enabling to Test Blockchain Sharding Protocols}

\author{Huawei~Huang\orcidA{}, ~\IEEEmembership{Senior Member,~IEEE}, Guang~Ye, Qinglin~Yang\orcidB{}, Qinde~Chen, Zhaokang~Yin, \\Xiaofei~Luo, Jianru~Lin, Jian~Zheng, Taotao~Li, 
Zibin~Zheng\orcidE{},~\IEEEmembership{Fellow,~IEEE}

\IEEEcompsocitemizethanks{
\IEEEcompsocthanksitem This is an updated version produced on Sep. 17, 2026.

\IEEEcompsocthanksitem This work was supported in part by the National Key R\&D Program of China under Grant 2022YFB2702304, in part by NSFC under Grant 62272496 and Grant 62032025, in part by the Fundamental Research Funds for the Central Universities, Sun Yat-sen University under Grant 23lgbj019, and in part by Tertiary Education Scientific research project of Guangzhou Municipal Education Bureau 2024312326. \textit{(Corresponding authors: Qinglin Yang; Xiaofei Luo.)}

\IEEEcompsocthanksitem Huawei Huang, Guang Ye, Qinde Chen, Zhaokang Yin, Xiaofei Luo, Jianru Lin, Jian Zheng, Taotao Li, and Zibin Zheng are with the School of Software Engineering, Sun Yat-Sen University, Zhuhai, Guangdong 519082, China (e-mail: huanghw28@mail.sysu.edu.cn; luoxf37@mail.sysu.edu.cn).

\IEEEcompsocthanksitem Qinglin Yang is with the Cyberspace Institute of Advanced Technology, Guangzhou University (Huangpu), Guangzhou, Guangdong 511363, China (e-mail: yangqinglin@gzhu.edu.cn).

\IEEEcompsocthanksitem Digital Object Identifier 10.1109/TSC.2025.3547222
}
}
\maketitle

\begin{abstract}

Numerous blockchain simulators have been proposed to allow researchers to simulate mainstream blockchains. However, we have not yet found a testbed that lets researchers develop and evaluate new consensus algorithms or protocols for blockchain sharding systems. To fill this gap, we developed BlockEmulator as an experimental platform, particularly for emulating blockchain sharding mechanisms. BlockEmulator adopts a lightweight blockchain architecture so developers can focus only on implementing their new protocols or mechanisms. Using BlockEmulator's layered modules and useful programming interfaces, researchers can implement a new protocol with minimal effort. In two steps, we test BlockEmulator's functionality. First, we prove the correctness of BlockEmulator's emulation results by comparing theoretical analysis with observed experimental results. Second, other experiments show that BlockEmulator can measure a range of metrics, including throughput, transaction confirmation latency, cross-shard transaction ratio, the queuing status of transaction pools, workload distribution across blockchain shards, etc. We have made BlockEmulator open-source on GitHub.
\end{abstract}


\section{Introduction}\label{sec:introduction}

Blockchain researchers need an experimental tool to verify the correctness or test the performance of their proposed protocols and algorithms. The literature has proposed a handful of blockchain simulators for different aims. For example, BLOCKBENCH~\cite{dinh2017blockbench} simulates and analyzes private blockchain performance. VIBES~\cite{stoykov2017vibes} enables blockchain simulation in large-scale network topologies and offers powerful visualization tools. SimBlock~\cite{aoki2019simblock} is an event-driven simulator suitable for studying the effect of block propagation in blockchain networks.
Other simulators (e.g., Ganache~\cite{mohanty2018deploying}, EthereumJS TestRPC~\cite{prusty2017building}, Parity Ethereum \cite{Open2023Fastest}, the built-in emulator of Hyperledger Fabric~\cite{androulaki2018hyperledger}, and Corda~\cite{brown2018corda}) focus on \hhw{simulating} the fundamental functions of a blockchain such as smart contracts and transactions.

In contrast, most experiments in existing blockchain publications are conducted through simulations. 
Their simulations must be implemented by building new blockchains~\cite{zamani2018rapidchain, al2017chainspace, dang2019towards, wang2019monoxide} or using existing blockchains~\cite{li2023cochain,li2023lbchain,hong2023gridb, kokoris2018omniledger}.
However, building a new blockchain from scratch requires significant effort and time, not to mention designing new protocols on top of it.
Existing blockchains like Ethereum~\cite{Wood2014Ethereum} have a vast codebase (\secondReply{approximately 220 thousand lines of code}), making them impractical for designing new consensus protocols. For instance, designing blockchain-sharding protocols on top of Ethereum is challenging, let alone redesigning other underlying layers, such as the storage and data layers.
In addition, when researchers seek to verify their protocols and algorithms, they aim to implement these solutions rapidly. They also need to initiate a testing platform and collect experimental data for performance evaluation.

\begin{figure}
    \centering
\includegraphics[width=0.48\textwidth]{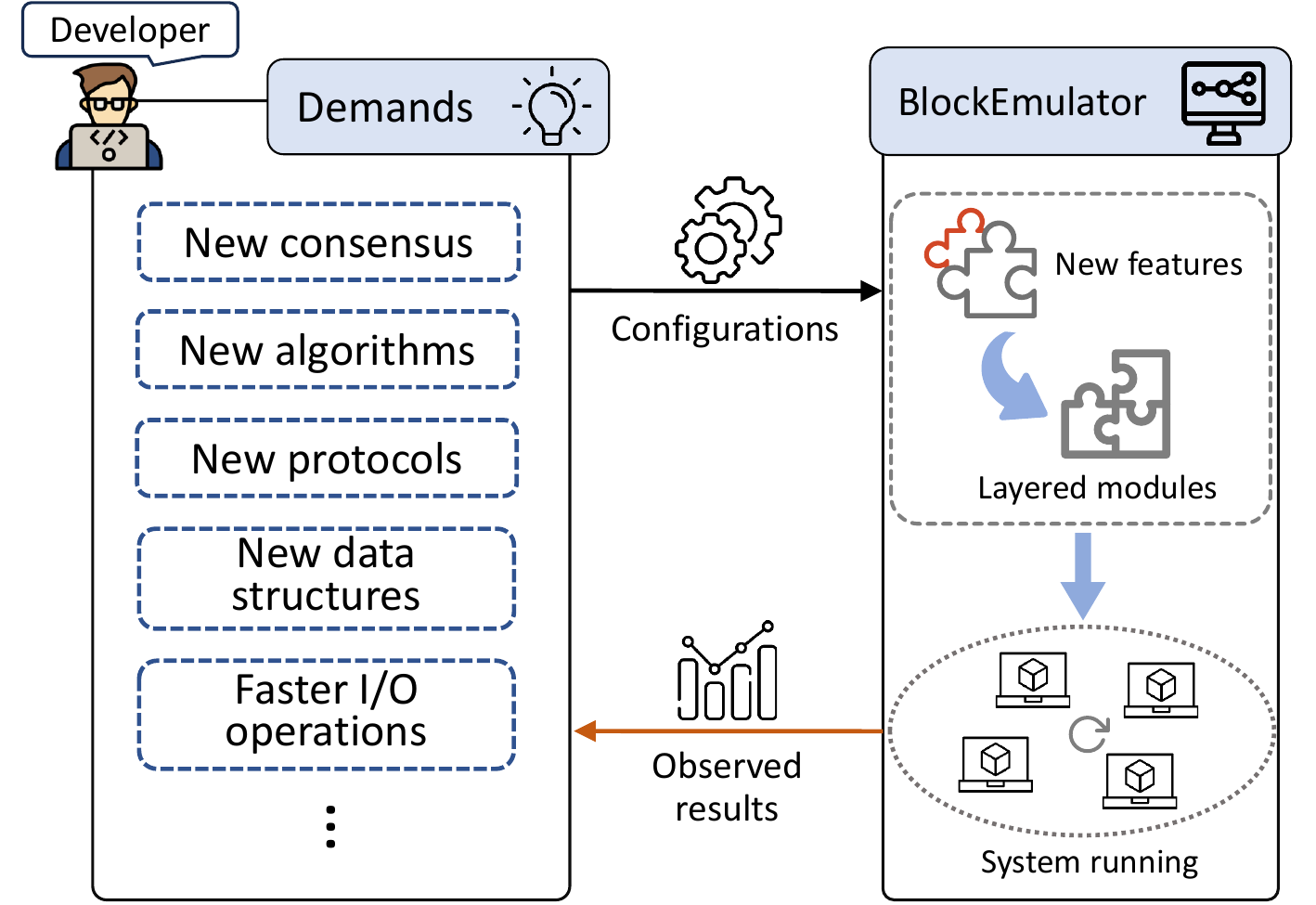}

    \caption{The design principle of \textit{BlockEmulator}.}
    \label{fig:DesignPrinciple}
\end{figure}

Although these existing open-source simulation codes \cite{li2023cochain,li2023lbchain,hong2023gridb} have their merits, they cannot meet the demands of verifying new consensus protocols for blockchain-sharding mechanisms. This is because blockchain sharding systems are complicated. It requires not only intra-shard consensus design but also shard-committee organization policies, inter-shard consensus algorithms, cross-shard communication mechanisms, security guarantees for blockchain shards, and atomicity for executing cross-shard transactions. Thus, researchers usually tailor blockchain-sharding mechanisms when designing a consensus algorithm for a sharded blockchain. When researchers change their research focus, they must implement another new blockchain sharding simulation framework as their experimental tool.

A better approach for blockchain researchers is to use a well-designed experimental platform that lets them add new sharding features with minimal effort.
To design an experimental tool that supports sharding mechanisms, it is necessary to determine the layered architecture of the sharded blockchain and develop the required components accordingly. 
Each component should also provide sufficient interfaces so users can design new mechanisms without worrying about cross-layer operations.
For example, S-Store~\cite{qi2022sstore} provides a dedicated data structure for migrating shard data. If deploying a blockchain on S-Store, the focus would be mainly on implementing the blockchain's underlying database interfaces and making minimal modifications to other layer data structures.
Similarly, to deploy asynchronous BFT consensus algorithms on blockchain shards using Dispersedledger~\cite{yang2022dispersedledger}, users would only need to modify the consensus layer within each shard. In addition, to deploy account redistribution algorithms~\cite{li2022achieving}, users only need to modify partial communication mechanisms among shards in Dispersedledger's consensus layer.
Therefore, implementing an experimental platform for blockchain sharding mechanisms involves multiple functionalities such as state management, cross-shard transactions, data availability, and inter-shard consensus. 
Hence, developing an effective experimental tool for blockchain sharding presents the following challenges.

\begin{itemize} 
    \item \textbf{Modular architecture.} The tool requires a modular architecture to facilitate development across various layers of the blockchain system. This design ensures both flexibility and adaptability for different use cases.
    
    \item \textbf{Interfaces for customized implementations.} The platform must provide diverse interfaces to support developing complex sharding consensus protocols. It must also enable the measurement of various performance metrics.
    
    \item \textbf{Result accuracy.} Ensuring the correctness of experimental results is critical, as it directly affects the validity of research findings and the tool's reliability. 
\end{itemize}

However, we have not yet found such an experimental tool in the literature.
To fill this gap, we design a lightweight experimental platform, \textit{BlockEmulator}, that enables researchers to develop and test new blockchain sharding protocols. 
As shown in Fig.~\ref{fig:DesignPrinciple}, BlockEmulator implements only the core functions of a blockchain system, including a node manager, consensus module, and on-chain transaction storage. For example, a client only needs to configure BlockEmulator's experimental parameters to match its design. After performing emulation, researchers can collect the experimental results from log files.

BlockEmulator also includes popular consensus protocols for sharded blockchains, such as Proof-of-Work (PoW) and Practical Byzantine Fault Tolerance (PBFT). In particular, BlockEmulator offers the system-level design and implementation of blockchain sharding mechanisms. BlockEmulator implements cross-shard transactions using two representative solutions: the \textit{transaction relaying mechanism} proposed by Monoxide~\cite{wang2019monoxide} and the \textit{BrokerChain} protocol~\cite{huang2022brokerchain}.

The paper's contributions are summarized as follows.

\begin{itemize}
    \item BlockEmulator is designed as an open-source experimental tool for evaluating blockchain sharding protocols. It provides a modular architecture and rich interfaces that enable researchers to quickly implement customized algorithms, protocols, and mechanisms. It also offers useful functions to help researchers collect various experimental metrics.
    
    \item We designed experiments to demonstrate BlockEmulator's correctness. Experimental results demonstrate that BlockEmulator can measure various metrics of a sharded blockchain, including transaction throughput, transaction confirmation latency, the queueing status of transaction pools, cross-shard transaction ratio, workload distribution across blockchain shards, etc.
    
    \item BlockEmulator has attracted an initial group of users from GitHub \cite{blockemulatorGitHub}.
    
\end{itemize}

The remainder of this paper is organized as follows. 
Section~\ref{sec:background} provides background on PBFT, Sharding Mechanisms, and the Testing Environments of existing blockchains. Section~\ref{sec:analysis} provides preliminaries and analysis of the cross-shard mechanism.
Section~\ref{sec:designFramework} presents the module design and framework of BlockEmulator with details. Section~\ref{sec:Experiments} conducts extensive experiments to verify BlockEmulator's performance. Finally, Section~\ref{sec:conclusion} concludes this work.

%%%%%%%%%% ==================
%%%%%%%%%% ==================
%%%%%%%%%% ==================
\section{Background and Preliminaries}\label{sec:background}

This section introduces the background and preliminaries for BlockEmulator, including sharding mechanisms and representative blockchain experiment tools.

%%%%%%%%%% ==================
\subsection{Blockchain Sharding}

Sharding divides a single blockchain into multiple parallel shards to improve scalability. Each blockchain shard maintains the account state for a subset of transactions. Thus, the sharding mechanism can improve the throughput of the entire blockchain system. 

Blockchain sharding can be classified into three categories, i.e., network sharding, transaction sharding, and state sharding. Among them, state sharding has become a research focus because of its low storage requirements~\cite{huang2022brokerchain,wang2019monoxide}. 
Following the static state sharding approach, if a proof-of-work (PoW) blockchain is segmented into $S>0$ shards, the blockchain's performance can be scaled out by a factor of $S$. However, the blockchain system's computational power is distributed across all shards. Such segmentation of consensus power reduces the security level to $1/S$. Malicious nodes can more easily attack a specific shard with their consensus power. This is because, by the pigeonhole principle~\cite {trybulec1990pigeon}, there will always be a shard with computational power less than or equal to $1/S$. 
Sacrificing significant security for higher throughput is unacceptable for a blockchain system. Therefore, designing security mechanisms for blockchain sharding systems is crucial.

In blockchain systems that adopt the account/balance model, Monoxide~\cite{wang2019monoxide} is a representative solution. The ``Chu-ko-nu Mining'' consensus proposed by Monoxide ensures that the blockchain system's security does not decrease as the number of shards increases. All shards in Monoxide perform the PoW consensus protocol simultaneously. Unlike static state sharding, each shard's PoW can be added to other shards' local chains as long as it meets the mining difficulty. This approach amplifies the effective computational power within each shard by a factor of $S$ (i.e., the number of shards). This feature ensures the system's security. Monoxide adopts an address-suffix partitioning policy, aiming to prevent malicious nodes from intentionally concentrating in a single shard.

In state sharding that adopts the account/balance-based transaction model, the blockchain system consists of a final committee and several worker shards. The final committee is also called the \textit{main} shard. The blockchain sharding system usually runs multiple mechanisms rather than relying on a single pure consensus mechanism. Committee members are periodically elected using more secure mechanisms, such as PoW, to prevent Sybil attacks~\cite{zhang2019double} and bribery attacks~\cite{gao2019power}. \textit{Worker} shards generally use high-throughput consensus mechanisms like PBFT for block generation to improve system throughput. 
The main shard handles configurations and adjustments that improve blockchain performance. For example, the main shard periodically reconfigures the blockchain sharding system. The performance of the sharded blockchain is mainly provided by \textit{worker} shards. During runtime, the blockchain sharding system advances to the next epoch when a specific cycle expires or after a predefined number of blocks are generated. A reconfiguration phase runs before the next epoch begins.

Experimental evaluation of sharded blockchains typically requires significant effort.
Some researchers choose to implement a new blockchain from scratch to realize the protocol prototypes they propose, such as RapidChain~\cite{zamani2018rapidchain}, Chainspace~\cite{al2017chainspace}, SGX Sharding~\cite{dang2019towards}, and Monoxide. 
In these works, researchers need to implement many components unrelated to their core design, such as underlying storage, network communication, and cryptographic authentication, which undoubtedly increases the workload.
Other works build upon existing blockchain platforms, such as CoChain~\cite{li2023cochain}, LB-Chain~\cite{li2023lbchain}, GriDB~\cite{hong2023gridb}, S-Store~\cite{qi2022sstore}, and OmniLedger~\cite{kokoris2018omniledger}. By modifying open-source codebases, these works can reduce some development efforts. However, these open-source codebases are often quite large. For example, Ethereum~\cite{Wood2014Ethereum} contains 220K lines of code, and Hyperledger Fabric~\cite{androulaki2018hyperledger} has up to 1 million lines. The large size of these codebases makes secondary development and modification more challenging, and it may also introduce more complexity and code coupling issues. Thus, both methods involve significant workload and complexity.

%%%%%%%%%% ==================
\subsection{Representative Blockchain Experiment Tools}

Through a thorough literature review, we found the following representative simulators, emulators, or platforms that blockchain researchers can adopt as experimental tools.

\begin{itemize}
    \item Ganache~\cite{mohanty2018deploying}, formerly called TestRPC, is a blockchain emulator developed by Truffle Suite. It provides a local blockchain environment that allows developers to simulate Ethereum networks. Ganache supports features like smart contract development, transaction simulation, and account management.

    \item EthereumJS TestRPC~\cite{prusty2017building} is a \textit{Node.js}-based emulator that provides a local Ethereum blockchain for testing and development purposes. It offers a lightweight and configurable environment for running test code and deploying smart contracts.

    \item Parity Ethereum~\cite{Open2023Fastest} provides a fully functional environment with mining capabilities, account management, and smart contract deployment. It lets developers create a local private blockchain to test Ethereum applications.

    \item Hyperledger Fabric~\cite{androulaki2018hyperledger} is a permissioned blockchain platform, which provides a built-in emulator for testing smart contracts and applications. It provides a local development environment that allows developers to simulate a blockchain network, create and interact with channels, deploy and invoke chaincodes, and test consensus mechanisms.

    \item Corda~\cite{brown2018corda} is a blockchain platform designed for enterprise and includes a mock network feature that lets developers emulate a Corda network for testing and development. The mock network provides a simulated environment for deploying and testing Corda contracts, flows, and transactions.
\end{itemize}

Given the existing experiment tools, these efforts mainly focus on smart-contract functions and transaction execution. Nevertheless, we have not found a dedicated emulator that supports blockchain-sharding mechanisms and handles cross-shard transactions.

%%%%%%%%%% ==================
%%%%%%%%%% ==================
%%%%%%%%%% ==================
\section{Principles of Blockchain Sharding}\label{sec:analysis}

In this section, we elaborate on the principles of blockchain sharding mechanisms, including the terminology, concepts, and critical technical issues of blockchain sharding.

%%%%%%%%%% ==================
\subsection{Cross-Shard Transactions}

\begin{figure}[t]
    \centering
\includegraphics[width=0.46\textwidth]{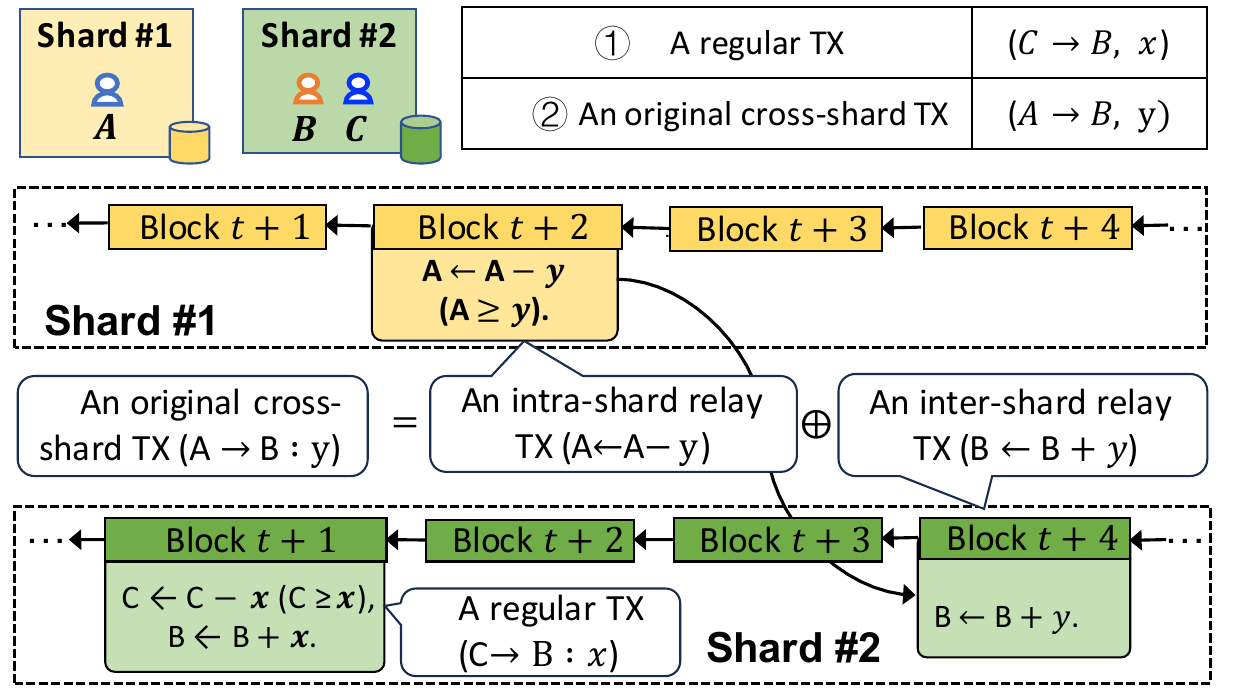}
    \caption{Illustration of multiple transaction types, including a regular transaction, an original cross-shard transaction, an intra-shard relay transaction, and an inter-shard relay transaction.}
    \label{fig:Tx_description}
\end{figure}

    As shown in Fig. \ref{fig:Tx_description}, suppose that there are three accounts $A$, $B$, and $C$, and two blockchain shards. Accounts $B$ and $C$ locate at Shard \#2, while Account $A$ locates at Shard \#1. Thus, the transaction ($C$ $\rightarrow$ $B$ : $x$) is called a \textit{regular} TX, which will be handled only in Shard \#2. In contrast, the transaction ($A$ $\rightarrow$ $B$ : $y$) becomes a \textit{cross-shard} TX.

    Monoxide~\cite{wang2019monoxide} divides an original cross-shard transaction ($A$ $\rightarrow$ $B$ : $y$) into an \textit{intra-shard} relay transaction ($A$ $\leftarrow$ $A$ - $y$) and an \textit{inter-shard} relay transaction ($B$ $\leftarrow$ $B$ + $y$), aiming to achieve the so-called \textit{eventual atomicity} of a cross-shard transaction.
    The intra-shard relay transaction deducts the funds from the payer's account in the source shard, while the inter-shard relay transaction deposits the funds into the payee's account in the destination shard. 
    These two relay transactions are connected through transaction-relaying messages.

    While handling all transactions in each shard, miners select transactions from the local transaction pool (abbr. as TX pool) to generate a new block. Miners can determine whether an original transaction in the block is a cross-shard transaction (abbr. as CTX) or an intra-shard regular transaction based on the locations of the payer and payee accounts.
    That is, the inter-shard relay transaction of the original CTX will be relayed to the payee's shard (i.e., the destination shard) after the block is committed on the payer's shard local chain.
    Once a consensus node receives an inter-shard relay transaction, it will check whether the related intra-shard relay transaction has been successfully included in the blockchain ledger. If the corresponding intra-shard relay transaction is on the chain, this consensus node adds the inter-shard relay transactions to its TX pool for future packaging.

    As shown in Table~\ref{tab:symbol&explanation} and Fig. \ref{fig:Tx_description}, to better understand the correlations between inside-shard and cross-shard transactions in a sharded blockchain, we let $\mathcal{U, V, W, X, Y, Z}$ denote the set of inter-shard relay transactions, intra-shard relay transactions, transactions packed into all historical blocks, all historical transactions submitted to TX pools by TX payers, all historical cross-shard transactions, and all historical regular transactions, respectively. Note that these non-cross-shard transactions are \textit{regular transactions}, which the local consensus within a shard can handle easily.
    Meanwhile, $|\mathcal{X}|$ represents the size of $\mathcal{X}$, and the same meaning applies to other symbols.

 Based on the aforementioned description, we have
    \begin{equation}\label{eq2}
        |\mathcal{Y}| = | \mathcal{U}| = |\mathcal{V}|.
    \end{equation}

A cross-shard transaction is divided into two relay transactions and will eventually be included in two blocks; we thus have the following equation.
\begin{equation}
\left\{\begin{array}{l}
    \left | \mathcal{Z} \right |+\left | \mathcal{Y} \right | =\left | \mathcal{X} \right |, \\
    \left | \mathcal{Z} \right | + 2 \cdot \left | \mathcal{Y} \right |=\left | \mathcal{W} \right |.
\end{array}\right. 
\end{equation}

As $|\mathcal{Y}|=|\mathcal{W}|-|\mathcal{X}|$ increases, the number of transactions packed in the blockchain will be significantly greater than the number of transactions processed in practice.
Thus, a sharding blockchain suffers from high latency and low throughput due to the cross-shard coordination.

\begin{table}[t]
    \centering
    \caption{Symbols defined for Transaction Types.}
    \begin{tabular}{m{0.8cm}<{\centering}|m{6.8cm}<{\centering}}
    \hline
    $\mathcal{U}$ & The set of all inter-shard relay TXs\\  \hline 
    $\mathcal{V}$ & The set of all intra-shard relay TXs\\ \hline 
     $\mathcal{W}$ & The set of all TXs \hhw{packed in historical blocks}\\ \hline 
     $\mathcal{X}$ & The set of all \hhw{historical TXs submitted to TX pools by \Guang{TX payers}}\\  \hline 
    $\mathcal{Y}$ & The set of all \hhw{historical} cross-shard TXs\\ \hline 
    $\mathcal{Z}$ & The set of all \hhw{historical} regular TXs\\  \hline 
    \end{tabular}
    \label{tab:symbol&explanation}
\end{table}

%%%%%%%%%% ==================
\subsection{Shard Workload Balance}\label{AcountRedistAlgorithm}

    Monoxide \cite{wang2019monoxide} allocates accounts to designated shards based on account suffixes.
    However, as shown in~\cite{huang2022brokerchain}, some accounts can be very active and thus create unbalanced workloads across shards. Monoxide neither considers how to reduce the proportion of CTXs nor how to balance shard workloads.

To the best of our knowledge, we have found two representative algorithms that can balance workload across blockchain shards: CLPA~\cite{li2022achieving} and BrokerChain\cite{huang2022brokerchain}. They are introduced as follows.

    CLPA~\cite{li2022achieving} is a dynamic account allocation algorithm that is different from the static policy defined in Monoxide\cite{wang2019monoxide}. It balances workloads across shards and reduces cross-shard transactions simultaneously.
    In~\cite{li2022achieving}, committee nodes are selected by PoW at the beginning of each epoch. 
    The committee then listens to newly generated blocks from each shard and builds an account graph from these blocks. In this graph, edges are the transactions in the blocks, and vertices are the accounts associated with those transactions. 
    When an epoch expires, the committee runs CLPA, which then returns the account reallocation results.
    The account graph is updated each epoch to balance workload across shards by invoking graph partition algorithms such as Metis or a community-aware graph partition algorithm.

Huang \textit{et al.}~\cite{huang2022brokerchain} proposed the BrokerChain protocol, aiming to reduce the number of cross-shard TXs. 
BrokerChain selects certain accounts as \textit{broker accounts}, each duplicated across multiple shards. 
In the BrokerChain protocol, a CTX is transferred into two intra-shard regular transactions. Those two regular TXs can be quickly handled in the payer's and payee's shards, respectively. A specific broker account \textit{bridges} the tokens transferred in the original CTX.
Note that if the payer or payee of a CTX is a broker account, this CTX becomes a regular transaction. That is why BrokerChain can significantly reduce the number of CTXs.

%%%%%%%%%% ==================
\subsection{The Composition of Sharding Consensus}

    In most blockchain sharding studies~\cite{luu2016secure, zamani2018rapidchain, huang2022brokerchain, li2022achieving}, a sharding consensus protocols are composed of \textit{worker} shards and \textit{main} shard (i.e., the final committee). 
    \textit{Worker} shards package blocks and achieve local consensus, while the \textit{main} shard runs reconfiguration algorithms (e.g., CLPA\cite{li2022achieving}) and feeds the results back to the \textit{worker} shards.

%%%%%%%%%% ==================
%%%%%%%%%% ==================
%%%%%%%%%% ==================
\section{Design and Implementation of BlockEmulator}\label{sec:designFramework}

This section elaborates on the design and implementation of BlockEmulator.

%%%%%%%%%% ==================
\subsection{Layered Architecture}

\begin{figure}[t]
    \centering
    \includegraphics[width=0.46\textwidth]{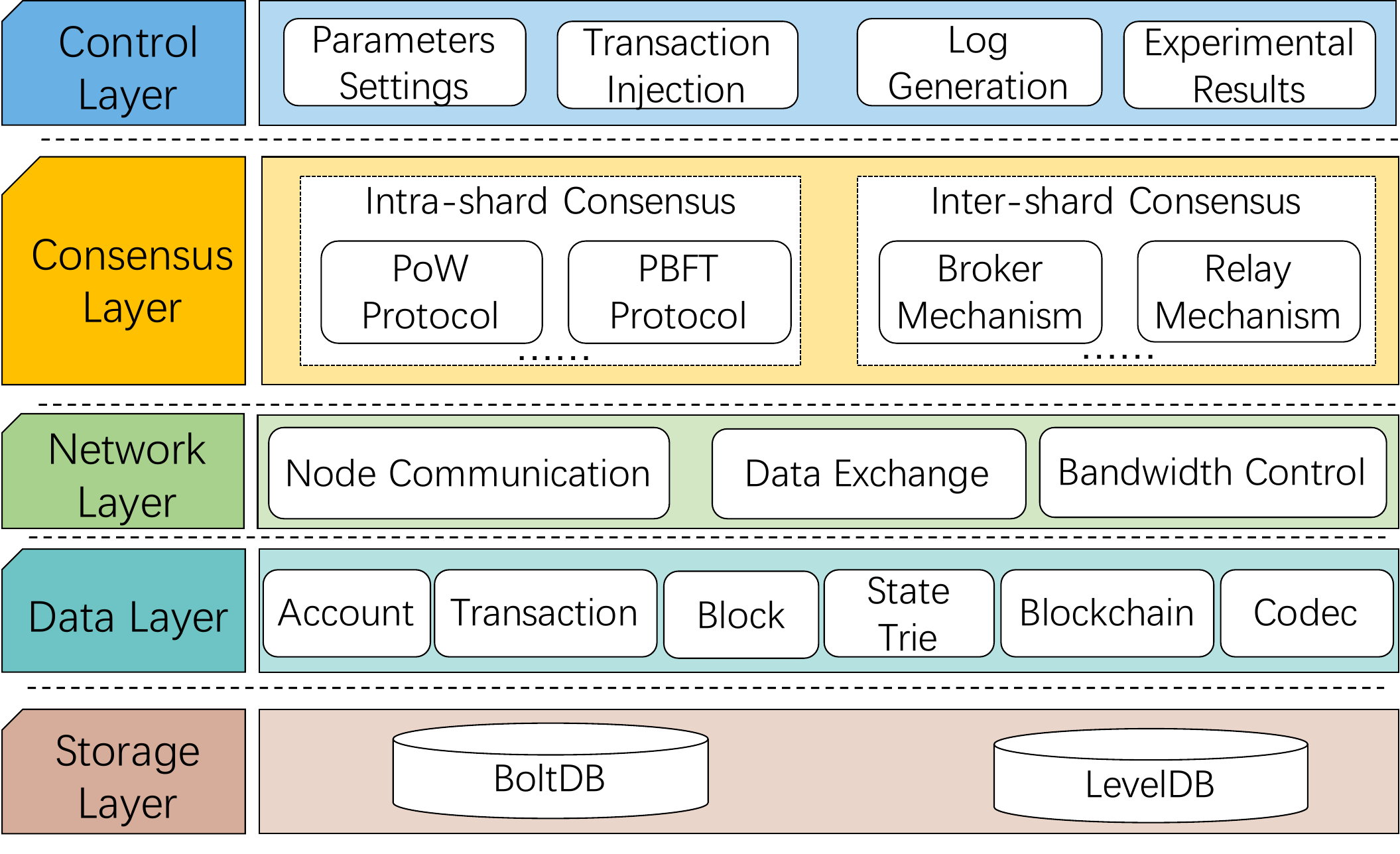}
    \caption{The layered architecture of \textit{BlockEmulator}.}  \label{fig:testbedArchitectures}
\end{figure}

Fig.~\ref{fig:testbedArchitectures} shows an overview of \textit{BlockEmulator} 's layered architecture, which consists of the \textit{Storage Layer}, \textit{Data Layer}, \textit{Network Layer}, \textit{Consensus Layer}, and \textit{Control Layer}.

%%%%%%%%%% ==================
\subsubsection{Storage Layer}

    The storage layer stores ledger data on disk, including generated blockchain data, experiment log files, etc. For instance, consensus nodes save the confirmed blocks, transactions, account states, and system logs.
    The storage layer uses boltDB and levelDB. In particular, we adopt the Merkle Patricia Tree (MPT) on top of levelDB for more efficient I/O operations.

%%%%%%%%%% ==================
\subsubsection{Data Layer}

 The data layer establishes the system's fundamental data structures, including accounts, transactions, blocks, state trees, and nodes.
 The data layer interacts with the storage layer. When running BlockEmulator, the data layer continually generates and encodes new blocks. These new blocks trigger the modifications of state trees. The system encodes these new tree nodes locally. 
 The encoded blocks and account states can be stored in local files by invoking the storage layer interfaces.
 Furthermore, after an experiment completes, the storage layer can retrieve the block data stored in files.
 The data layer can then decode the retrieved block data back into blocks. Similarly, the retrieved root nodes of state trees in different blocks can be used to generate the state trees locally.
 This way enables researchers to access the generated blocks and account state trees. Thus, users can examine account states offline to verify the correctness of transaction execution results produced by BlockEmulator.

%%%%%%%%%% ==================
\subsubsection{Network Layer}

The network layer supports communication across consensus nodes deployed on diverse devices.
TCP implements an end-to-end packet transmission mechanism to exchange data among nodes. The data exchanged includes transactions, blocks, consensus messages, etc.
We implement TCP as a keep-alive connection to support substantial message exchanges within the network. 
In this manner, the TCP connection between two nodes is initiated and closed only once during the overall emulation.
%

%%%%%%%%%% ==================
\subsubsection{Consensus Layer}

The consensus layer enables consensus nodes to agree on a new block.
In a sharded blockchain, nodes run both the intra-shard and inter-shard consensus mechanisms. 
The intra-shard consensus mechanisms (like PoW and PBFT) ensure transaction consensus within a shard. 
The inter-shard mechanisms guarantee communication across shards. For example, to handle a CTX, nodes in the payer's and payee's shards should receive messages from other shards.

%%%%%%%%%% ==================
\subsubsection{Control Layer}

The control layer manages system execution. 
Before an experiment starts, this layer must prepare the experiment environment.
Preparing the experiment environment includes initializing metric-testing threads, starting to listen for messages, and determining each consensus node's host.
When BlockEmulator runs, the control layer injects transactions into each shard's transaction pool at a predefined rate. You can retrieve the transactions injected into BlockEmulator from the historical transactions of any blockchain, such as Ethereum.
This layer also records information about blocks generated by shards, including not only the blocks themselves but also the necessary shard details.
After BlockEmulator finishes an experiment, users can collect experimental results from logs for scientific analysis.

\begin{figure}[t]
\centering
\includegraphics[width=\linewidth]{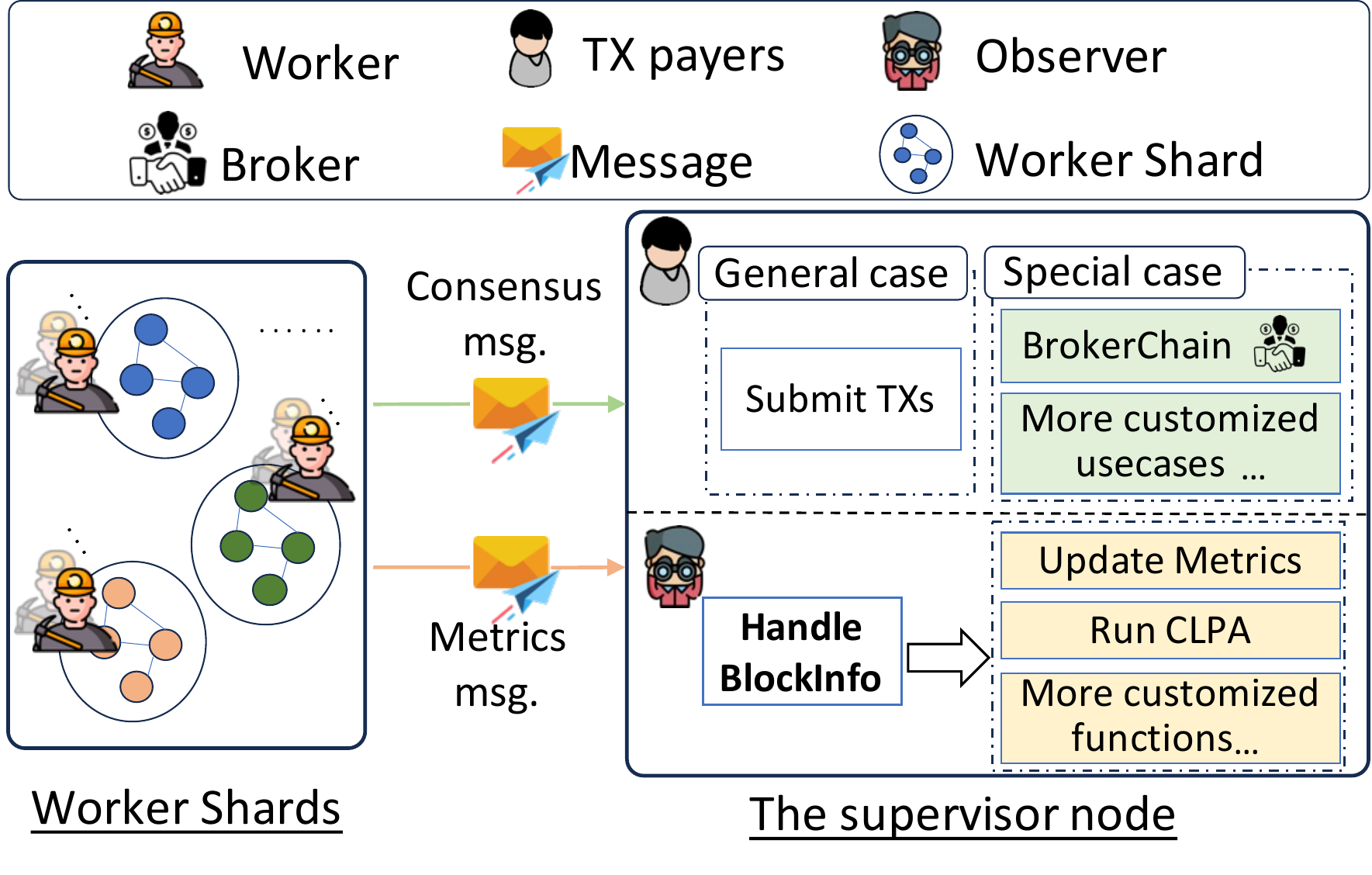}
\caption{Major roles and modules of BlockEmulator, and the interactions between \textit{worker shards} and the \textit{supervisor} node.}
\label{fig:SupervisorInterface}
\end{figure}

%%%%%%%%%% ==================
\subsection{Major Roles in BlockEmulator}\label{sec:roleOfBlockEmulator}

As shown in Fig. \ref{fig:SupervisorInterface}, \textit{BlockEmulator} has 2 node types: \textit{worker} nodes and \textit{supervisor} nodes. 
A \textit{worker} packs transactions and generates blocks for the blockchain in each local shard. 
\textit{Supervisor} nodes conduct global control behaviors, e.g., injecting transactions to shards, collecting committed blocks, and calculating system metrics. The number of required worker and supervisor nodes depends on the blockchain network scale. Developers can configure the number of these two node types flexibly according to their budget.

\textit{Supervisors} execute behaviors concurrently. Different \textit{supervisors} could commit different behaviors collaboratively. However, this multiple-supervisor implementation adds significant complexity to the emulation system. To simplify \textit{BlockEmulator}, we implement two roles, i.e., \textit{TX payers} and \textit{observer}. The first role injects transactions into each shard's TX pool. 
As the second role,  \textit{observer} not only monitors the execution status of the blockchain system but also computes and records experimental outcomes.

In real blockchains, transaction payers initiate transactions and observers collect blocks. Accordingly, we model these roles and add additional features in our BlockEmulator. 
Both the \textit{TX payers} and the \textit{observer} can extend functionality by invoking the interfaces presented in Section~\ref{sec:CoreInterfaceModules}.

%%%%%%%%%% ==================
\subsection{Core Interface Modules}\label{sec:CoreInterfaceModules}

\begin{figure}[t]
    \centering
    \includegraphics[width=0.48\textwidth]{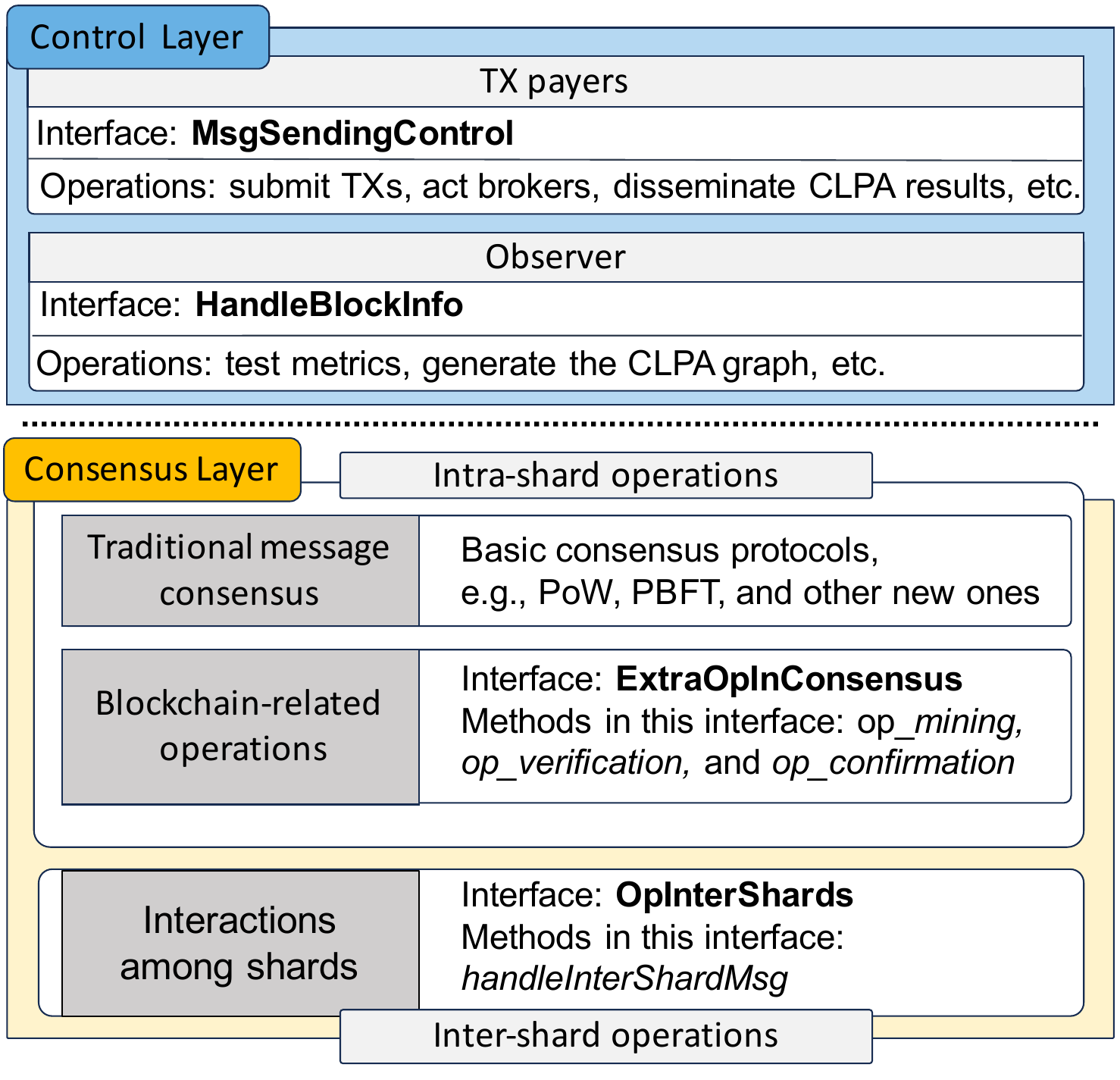}
    \caption{The interfaces designed for Consensus Layer and Control Layer.}  \label{fig:interfacesofTwoLayers}
\end{figure}

To ease secondary development with BlockEmulator, we implement popular blockchain mechanisms as pre-installed modules and design application programming interfaces (APIs) to ensure compatibility among mechanisms. We show the primary APIs in Fig.~\ref{fig:interfacesofTwoLayers}.
We believe users can easily implement new mechanisms in a few lines of code using the interfaces we provide.

%%%%%%%%%% =========
\subsubsection{The Intra-shard Interface in Consensus Layer}\label{subsubsec:intra-shard-interface}

In general, blockchain consensus operations fall into two categories: \textit{traditional message consensus} and \textit{blockchain-related operations}.

For example, the PoW mechanism predates blockchain and was used to protect against spam emails~\cite{dwork2005pebbling}. In other words, PoW is a traditional message-consensus mechanism that can be used in other distributed systems. 
However, in a blockchain environment, PoW must also perform blockchain-related operations, such as packing transactions, verifying the correctness of all transactions in a block, and adding a new block to the blockchain.

Likewise, by incorporating blockchain-related operations, the traditional PBFT protocol can work in blockchain systems. 
Therefore, the details of these operations may differ across blockchains. Users should focus on their needs when designing new mechanisms on top of \textit{BlockEmulator}. The traditional message-consensus part can be reused because it does not need modification in most scenarios.

As shown in the analysis depicted above, we separate the \textit{traditional message consensus} from \textit{blockchain-related operations} in our implementation.
We first design the interface \textbf{ExtraOpInConsensus} within the consensus layer to enable \textit{blockchain-related operations}. \hw{These interfaces are then named according to their relevant functionalities}, including \texttt{op\_mining}, \texttt{op\_verification}, and \texttt{op\_confirmation}. Here, \texttt{op\_mining} occurs during the message-generation phase, \texttt{op\_verification} during the consensus phase, and \texttt{op\_confirmation} when consensus is achieved.

For instance, to deploy an account-migration protocol in \textit{BlockEmulator}, you can implement the \textit{ExtraOpInConsensus} interface in the following three steps.
Step i). In the account-migration protocol, the associated account states are transferred from one shard to another. To facilitate this process, during the \textit{mining} phase, the block generator proposes either a normal block containing only TXs or a special block encompassing the account states to be migrated. Thus, the method \texttt{op\_mining} should be re-implemented to support two block types. 
Step ii). When receiving the special block, worker nodes should validate the proofs of the account states in the block. 
%通过更新这个 op_verification 函数，让它包含 verification 这个功能，可以让 worker 节点准确地 check 
Worker nodes can validate accurately by updating the function \texttt{op\_verification}.
Step iii). When a specific block is committed, all account states contained in the block are migrated. Then, worker nodes must update the state trie with the latest migrated account states. Consequently, the method \texttt{op\_confirmation} should be modified to achieve this functionality.

With these interface modifications to \textit{ExtraOpInConsensus}, worker nodes can effectively handle intra-shard consensus for account migration.

\subsubsection{The Inter-shard Interface in Consensus Layer}\label{subsubsec:inter-shard-interface}

In a sharded blockchain, a shard must interact with other shards to synchronize some necessary information. However, the traditional consensus, such as PoW and PBFT, do not provide such functionality. 
To solve this problem while enabling secondary development, we define the \textbf{OpInterShards} interface, which helps users define message-handling rules and broadcast them across shards.

We continue to use account migration to implement the \textit{OpInterShards} interface, as follows.
Step i). Worker nodes process the account partition messages that contain information about those accounts designated for migration to different shards. Worker nodes retrieve the relevant account states from their state tries, consolidate these states into account-migration messages, and send these messages to their corresponding destination shards. By implementing \texttt{handleInterShardMsg} in the \textit{OpInterShards} interface, worker nodes can send account-migration messages accurately.
Step ii). When worker nodes receive the account-migration message, they determine whether the account states in the message are destined for their shards. Once worker nodes collect all required account states, they propose a special block, as detailed in Sec.~\ref{subsubsec:intra-shard-interface}. To ensure accurate account migration, this interface must implement message handling and trigger functions effectively.

%%%%%%%%=========
\subsubsection{Interfaces in Control Layer} 

Recall that, in Section~\ref{sec:roleOfBlockEmulator}, the \textit{supervisor} nodes play two roles (\textit{TX payers} and \textit{observer}) in the control layer. %
In practice, these two roles can be merged in only one \textit{supervisor}, we thus implement one \textit{supervisor} in the current version of BlockEmulator. 

As illustrated in Fig.~\ref{fig:SupervisorInterface} and Fig.~\ref{fig:interfacesofTwoLayers}, we present two major \textit{supervisor} interfaces. With these 2 interfaces, users can extend the \textit{supervisor} with more features to meet their needs. We explain the functions of these two interfaces as follows.

\textbf{MsgSendingControl Interface}. 
Although the primary role of \textit{TX payers} is to submit transactions, the \textit{MsgSendingControl} interface lets users extend their application purposes.
For example, when an account redistribution algorithm runs on the blockchain, \textit{TX payers} can extend their functionality by rewriting the \textit{MsgSendingControl} interface. This rewriting lets \textit{TX payers} disseminate updated account-partition results to worker shards.
For another example, when users implement the BrokerChain protocol~\cite{huang2022brokerchain}, \textit{TX payers} can play the \textit{broker} account's role as the transaction bridge between shards.
%
%%\reviseyg{Exactly, \textit{TX payers} execute additional operations, including partitioning a raw CTX into two corresponding broker TXs and submitting the broker TXs once the relevant ones have been committed.}
%
The \textit{MsgSendingControl} interface enables the \textit{supervisor} node to proactively transmit messages, driven by data inputs such as TXs and account partition results. This is why we call the role associated with this interface the \textit{TX payer}. This is because a TX payer acts as an active request sender in a real-world blockchain network.

\textbf{HandleBlockInfo Interface}. Unlike the proactive \textit{TX payer}, the \textit{observer} analyzes received block information from consensus nodes. With the dedicated \textit{HandleBlockInfo} interface, the \textit{observer} can use the block information. 
For instance, when \textit{BlockEmulator} emulates an account redistribution algorithm, \textit{observer} needs to generate an account graph using the TXs collected from the previous epoch. It then uses this account graph to generate account-partition results.

Users can also define how to measure their metrics using the \textit{HandleBlockInfo} interface.
By implementing \textit{HandleBlockInfo}, users can define custom logic to measure their metrics of interest.
To measure metrics, redefine three functions in the \textit{HandleBlockInfo} interface: a \texttt{constructor} function for the data structure, a \texttt{metric-updating} function, and a \texttt{result-output} function. At system initialization, the system creates a dedicated data structure specified by this interface. This data structure processes and stores metric-related information. When it receives block messages from worker nodes, the \textit{observer} automatically updates this data structure using the \texttt{metric-updating} function. When the system terminates, the \textit{observer} invokes the result-output function to compute the final metrics from the data structure and then writes the results to disk. Thus, this customizable interface lets users collect and manage experimental results conveniently.

%%%%%%%%%% ==================

\subsection{Other Featured Implementations}

\subsubsection{Implementation of Account Redistribution}

In Section~\ref{AcountRedistAlgorithm}, we discussed CLPA algorithm~\cite{li2022achieving}. The \textit{Main} shard generates an account graph from blocks collected during an epoch and runs CLPA on the graph when the epoch expires.

We implement CLPA in \textit{BlockEmulator} and separate the graph-related functions from the consensus layer. Users can thus invoke CLPA even in a non-blockchain environment. 
We also prune CLPA results to reduce their sizes. Accounts are called \textit{dirty} if they need to migrate, and only these accounts are included in the account redistribution result.  

To reduce the communication overhead among nodes, \textit{BlockEmulator} recruits \textit{observer} nodes as the members of the \textit{Main} shard. 
In our demonstration code, we implement the CLPA graph-generation function and the CLPA redistribution algorithm in the \textit{HandleBlockInfo} and \textit{MsgSendingControl} interfaces, respectively. The interface \textit{HandleBlockInfo} will be triggered by the block information message sent from the \textit{workers}. This mechanism supports ongoing CLPA graph generation based on newly arrived blocks. When reconfiguration conditions are met, e.g., after a predetermined time interval or a specified number of transactions have been processed, the interface \textit{MsgSendingControl} is invoked. This triggers the CLPA redistribution algorithm, and the results are then disseminated to \textit{worker} shards.

%%%%%%%%%% ==================
\subsubsection{Implementation of Account Migration}

When \textit{worker} shards receive the account redistribution result, they need to save it and perform account migration. 
To enable this, rewrite the \textit{ExtraOpInConsensus} and \textit{OpInterShards} interfaces carefully.

\textit{BlockEmulator} adopts a lock-based mechanism to implement the account migration.
Once the leader node of a \textit{worker} shard receives the account redistribution command, it first completes the current round of PBFT, then locks the associated account states in this shard. After these preparations, the leader node will handle the account-migration command. Thus, the interface \textit{ExtraOpInConsensus} should handle the message containing the account redistribution result.
By implementing this interface, the leader node determines which account information to send to the newly designated shard. The information includes account states and the transactions associated with \textit{dirty} accounts.

%如果 这个分片收集到了所有来自其他分片的 dirty account 的信息，那么它就将 该分片的 account states 解锁。然后这个分片会进行关于账户迁移的共识。通过修改 opintershard 接口，这种共识将会被无缝地实现。按具体的实现来说，用户可以为账户迁移定义一种新的 区块类型，并且在 接口内的 各个 methods 中加入相关的处理逻辑。这种方法可以保证代码的结构性，而且有利于自定义方法。

A shard unlocks its account states when it receives all \textit{dirty} account information from other shards. Then, this shard executes the account-migration consensus, which can be achieved by modifying the \textit{OpInterShards} interface. Users can define a new block type dedicated to account migration. This approach ensures codebase compatibility while enabling custom functionality.
Users can also replace the account migration module to meet their needs.

The examples above show that users can customize their protocols or algorithms using the provided interfaces. However, users should keep the following reminders in mind.
First, users must clearly define which nodes execute what functions, and where to implement their customized functionalities in either a \textit{worker} or a \textit{supervisor} node.
Second, users should determine the timing of function execution, which affects which interfaces need to be rewritten.
Finally, users need to implement the core components of their customized protocol.
As an application instance for account migration in the context of a sharding blockchain, please refer to our latest work \cite{huang2024account} for more details. The implementation code is available in the \textit{Fine-tune-lock} branch of BlockEmulator's codebase \cite{blockemulatorGitHub}.

%%%%%%%%%% ==================
\subsection{Workflow of BlockEmulator}\label{subsec:workflow}

\begin{figure}[t]
    \centering
\includegraphics[width=0.48\textwidth]{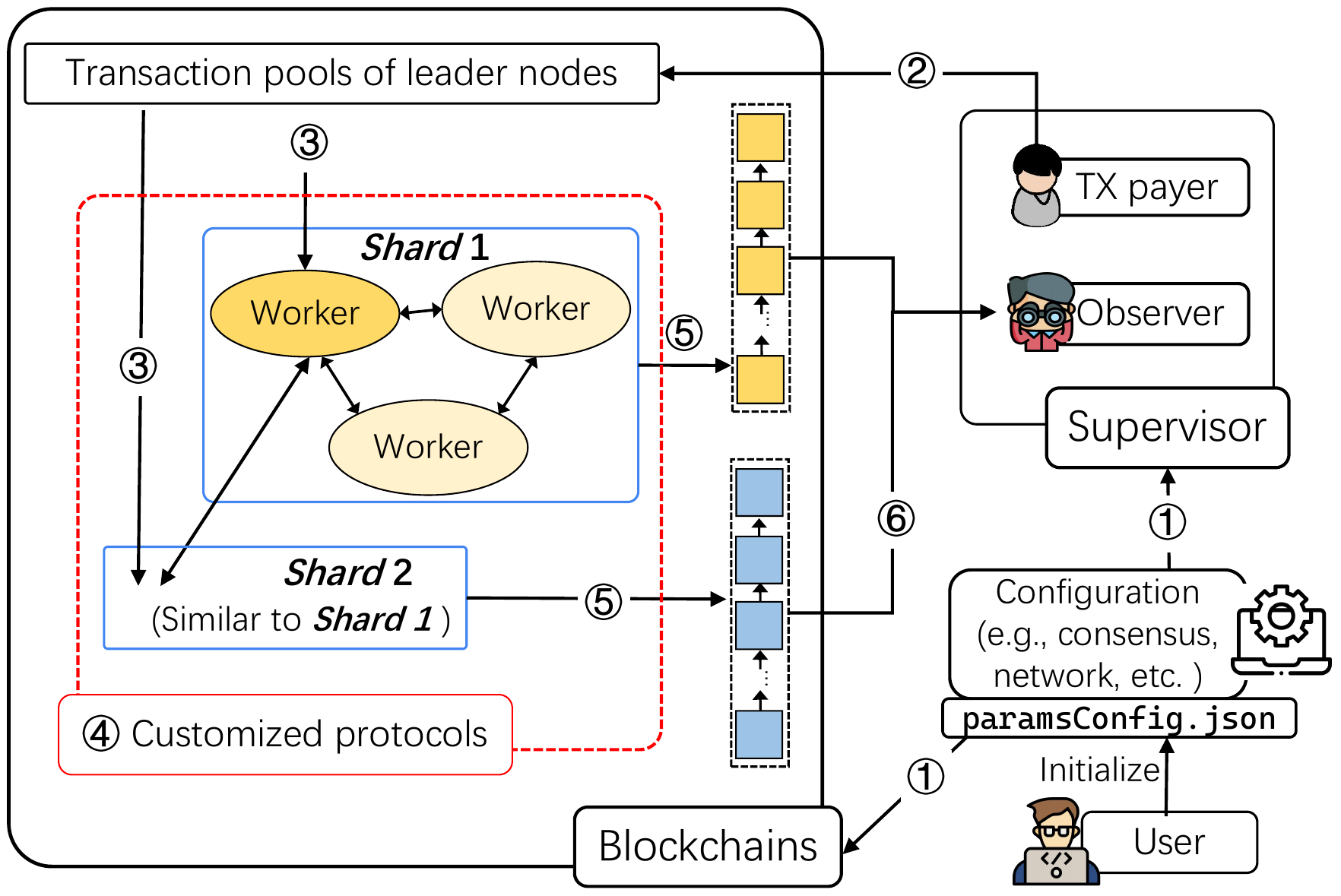}
    \caption{BlockEmulator workflow. The six steps for using BlockEmulator are as follows: \circled{1}: Developers set up the blockchain configuration and the supervisor by configuring parameters in \texttt{paramsConfig.json}. \circled{2}: The supervisor injects transactions. \circled{3}: Leaders load transactions from the pool. \circled{4}: Worker nodes operate according to customized protocols. \circled{5}: All worker nodes commit blocks on the chain, update the state tree, and store data locally. \circled{6}: Worker nodes send block information to the supervisor.}
    \label{fig:workflow}
\end{figure}

 \textit{BlockEmulator} works following the workflow (i.e., Fig.~\ref{fig:workflow}), where we take PBFT as an illustrative intra-shard consensus.
 
 \begin{itemize}
     \item In step \circled{1}, before the BlockEmulator system starts, users/developers set the configuration parameters for the emulated blockchain.
     The configuration parameters mainly include the output directory for experimental results, the block-generation interval, the adopted consensus protocol, block size, the number of nodes per shard, and the total number of blockchain shards across the network. Then, the BlockEmulator system generates a blockchain according to the configured parameters.
     When using PBFT, a \textit{leader} node is elected for each shard, followed by initializing all \textit{worker} nodes.
    
     \item In step \circled{2},
     the \textit{supervisor} node simulates the role of \textit{TX payers}. It continuously loads transactions from a local dataset and injects them into each shard's TX pool at a predefined static or dynamic rate. The system makes both the dataset directory and the transaction injection rate configurable parameters.

     \item Next, in step \circled{3}, the \textit{leader} node of each shard begins managing a set of incoming transactions received from the \textit{supervisor}. The \textit{leader} selects these transactions for inclusion in the newly generated block. The \textit{leader} uses a predefined transaction-selection mechanism, such as a strategy that maximizes fee revenue by prioritizing transactions by fee, or a first-in, first-out (FIFO) ordering policy.
     
     \item In step \circled{4}, \textit{worker} nodes operate according to user-specified protocols, including both inner-shard and inter-shard protocols. Nodes invoke the methods defined in the \textit{ExtraOpInConsensus} interface when handling inner-shard consensus. Taking PBFT as an example, after the \textit{leader} node selects a set of transactions and packages them into a new block, the method \texttt{op\_mining} in the \textit{ExtraOpInConsensus} interface will be invoked (in the \textit{pre-prepare} phase). This block is considered a PBFT proposal and is then proposed by the \textit{leader} node within its shard. 
     During the PBFT consensus, the method \texttt{op\_verification} is triggered when a node receives a block from another node (e.g., in the \textit{prepare} phase of PBFT). The method \texttt{op\_confirmation} is then invoked when a node commits a block (e.g., in the \textit{commit} phase of PBFT). These methods build the consensus logic for each shard's blockchain. They mitigate the risk of erroneous or malicious blocks being committed to the ledger.

    \item When an operation requires information from other shards, nodes communicate across shards. 
    This communication supports inter-shard events such as cross-shard transaction verification, cross-shard account migrations, and transaction-proof deliveries. 
    As shown in Fig.~\ref{fig:interfacesofTwoLayers}, the method \texttt{handleInterShardMsg} is defined in the \textit{OpInterShards} interface.
    Through customized protocols, users can define a unique message-handling logic in this method. Upon receiving a message, a \textit{worker} node extracts the message header from the raw message and delivers it to the appropriate function. 
    For instance, when a node receives a CTX message, it determines the message type and routes it to the appropriate function responsible for handling CTX messages. This function validates the CTXs and appends them to the TX pool.

    \item When \textit{worker} nodes reach consensus through PBFT in step \circled{4}, each node appends the block to the blockchain in step \circled{5}. During step \circled{5}, the nodes also update the state trie and persistently store the block on local disk.
     
    \item Throughout the execution of the emulated blockchain, the \textit{supervisor} serves as the role of an \textit{observer}. It continuously monitors the operations and information across the entire blockchain and updates metric data as outlined in step \circled{6}. By implementing the interface \textit{HandleBlockInfo}, the \textit{supervisor} aggregates shard information from all shards. This allows the \textit{supervisor} to analyze the system's performance from multiple aspects, including transactions per second, transaction confirmation latency, transaction distributions, etc.

 \end{itemize}

 The supervisor stores the metric data and block information it collects on the local disk for experimental analysis.

%%%%%%%%%% ==================
%%%%%%%%%% ==================
%%%%%%%%%% ==================
\section{Experiments and Analysis}\label{sec:Experiments}

    This section presents comprehensive measurements and tests to show that BlockEmulator is useful and produces correct emulation results.

%%%%%%%%%% ==================
\subsection{Performance Metrics to Verify BlockEmulator}

When analyzing the functional performance of \textit{BlockEmulator}, we mainly focus on four metrics, including \textit{transactions per second} (TPS)~\cite{king2012ppcoin}, \textit{transaction confirmation latency} (TCL)~\cite{huang2022brokerchain}, \textit{cross-shard transaction ratio} (CTX Ratio)~\cite{zhang2023txallo} , and \textit{TX pool size}. A more detailed description of each metric is as follows. 

\begin{itemize}
    \item  \textbf{TPS} measures a blockchain system's transaction-handling capacity. A cross-shard transaction is deemed confirmed if its \textit{intra-shard} and \textit{inter-shard} relay transactions are both on-chain. Thus, we count an intra-shard relay transaction or an inter-shard relay transaction as 0.5 in our TPS calculation.

    \item \textbf{TCL} is the interval from the timeslot when a transaction is proposed to the timeslot when it is confirmed. In the context of blockchain, it refers to the time consumed between when a transaction is broadcast to the network and when it is included in a block and confirmed by the network.
    
    \item \textbf{CTX Ratio} refers to the proportion of transactions whose payer and payee accounts are located in different shards out of the total number of transactions submitted to TX pools. A transaction requires less overhead for message propagation and storage if its payer and payee accounts are located in the same shard. A higher CTX Ratio might lead to longer latency and lower throughput.
    
    \item \textbf{TX pool size} stands for the length of the TX pool queue. It can reflect real-time changes in the TX pool and indicate each shard's transaction workload.
\end{itemize}

%%%%%%%%%% ==================
\begin{table}[t]
    \centering
    \caption{Symbols used in experimental settings.}
    \begin{tabular}{m{1.6cm}<{\centering}|m{5.7cm}<{\centering}}
    \hline
    $\Theta$  &  Size of a block\\ \hline
    $\delta$      & Interval between two blocks (second)\\ \hline
    $\mathcal{R}$ & The set of transactions handled \\  \hline 
    $N$ & Number of shards \\ \hline 
    $t$ & Time cost in the phase \\ \hline
    $\phi$ & The speed of transaction packing \\ \hline
    $\mathbb{E}^\prime(\text{TPS})$ & TPS expectation for a single shard\\ \hline 
    $\mathbb{E}(\text{TPS})$ & TPS expectation for the whole system\\ \hline
    $\text{TCL}_\mathcal{Z}$ & TCL expectation for regular TXs\\ \hline 
    $\text{TCL}_\mathcal{Y}$ & TCL expectation for cross-shard TXs\\ \hline 
    
    \end{tabular}
    \label{tab:symbolinExperiments}
\end{table}
%%%%%%%%%% ==================

%%%%%%%%%% ==================
\subsection{Experimental Settings}

    In the following experiments, a block can maintain a maximum of 2,000 transactions.
    The interval between two blocks in each shard is set to 8 seconds. Each epoch lasts for 80 seconds. 
    These settings allow users to analyze the system's performance in different epochs.
    The experiments are conducted on 10 resource-heterogeneous physical machines running Ubuntu 20.04.6 LTS. 
    The hardware configurations of those physical machines are 5 Xeon(R) W-2150B CPUs and 64 GB RAM, 3 Intel (R) Core(TM) i7-9700F CPUs and 64GB RAM, 1 Intel (R) Core(TM) i7-9700F CPU and 16 GB RAM, and 1 AMD Ryzen 9 3900x 12-core CPU and 16 GB RAM.
    Moreover, we deploy multiple shards on each physical machine. Most \textit{BlockEmulator} modules are implemented in Go. Please find the open-source implementation from our GitHub project \cite{blockemulatorGitHub}.

    \textbf{Datasets.} The experiments utilize both a randomly generated synthetic dataset and a historical Ethereum dataset. 
    \textit{Dataset i)} The synthetic dataset, comprising load-balanced transactions, verifies the correctness of \textit{BlockEmulator}. 
    \textit{Dataset ii)} The historical dataset, sourced from the Ethereum blockchain~\cite{Zheng2020xblock}, includes real-world transactions (excluding smart contract interactions) from block heights 2,000,000 to 2,999,999.

    \textbf{Launching BlockEmulator.} By modifying parameter configurations and IP-address files, we use a pre-compiled executable to generate batch command files for each shard. Next, we can launch all nodes to run BlockEmulator. The BlockEmulator launch guide is available in our GitHub project \cite{blockemulatorGitHub}.

%%%%%%%%%% ==================
\subsection{Correctness Verification of BlockEmulator's Observed Results}

To demonstrate the correctness of \textit{BlockEmulator}, we conduct a large-scale set of experiments. We evaluate the TPS and TCL versus different numbers of shards. We also set the total number of transactions $\mathcal{W}$ to $100,000 \times N$, where $N$ is the number of shards in each experiment group. Thus, the system runs long enough under different settings of $N$.

For a more detailed validation, we first utilize a manual dataset to generate load-balanced transactions randomly across all shards. We also configure transaction fees so that transaction priorities are equal in TX pools. Each shard's TX pool thus becomes a FIFO queue. 
To mitigate the impact of transaction injection, the nodes collect all transactions and put them into TX pools before the consensus starts. In addition, the system adopts the \textit{Relay} mechanism~\cite{wang2019monoxide} to process cross-shard transactions and applies a static sharding algorithm. 
Finally, we calculate the expected TPS performance across all 3 phases of transaction handling and analyze the TCL distributions for both regular and cross-shard transactions. We then compare the experimental results obtained from \textit{BlockEmulator} with the theoretical computations.

\begin{figure}[t]
\centering
\includegraphics[width=0.48\textwidth]{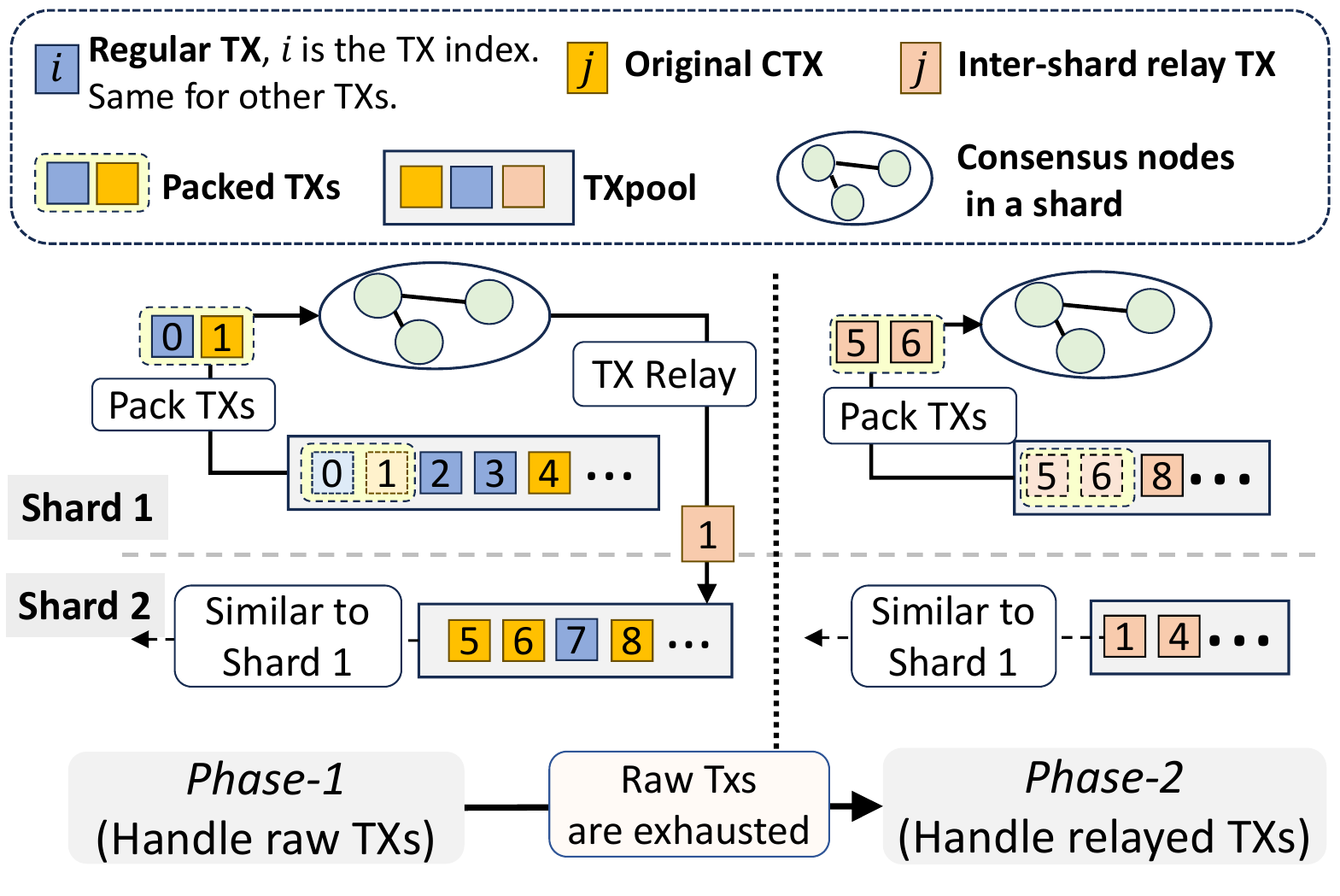}
\caption{Correctness demonstration of the TX pools in Phase-1 and Phase-2. These two phases are defined to help explain how TX pools evolve when given a group of injected raw transactions.}
\label{fig:phaseEvolutions}
\end{figure}

\begin{figure*}[t]
\centering
\includegraphics[width=\textwidth]{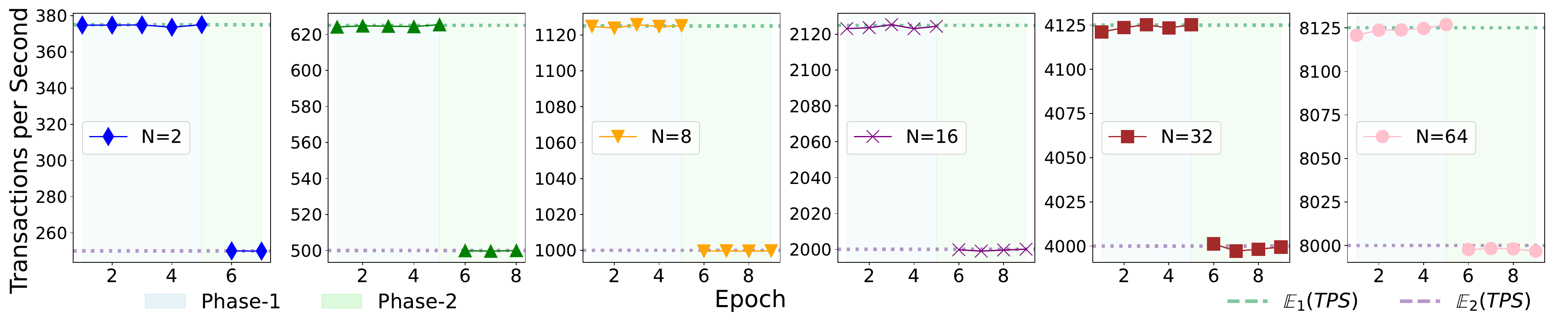}
\caption{Comparison of the theoretical and experimental results in terms of TPS. The two horizontal dashed lines in each figure represent the $\mathbb{E}_{1}(\text{TPS})$ and $\mathbb{E}_{2}(\text{TPS})$ in the corresponding setting of shard numbers.}
\label{fig:exp2-TPS_allShardNumber}
\end{figure*}

\begin{figure*}[t]
\centering
\subfigure[The proximity (measured in percentage) between the experimental results and expectations.]{
\begin{minipage}[t]{0.27\linewidth}
\centering
\includegraphics[width=\textwidth]{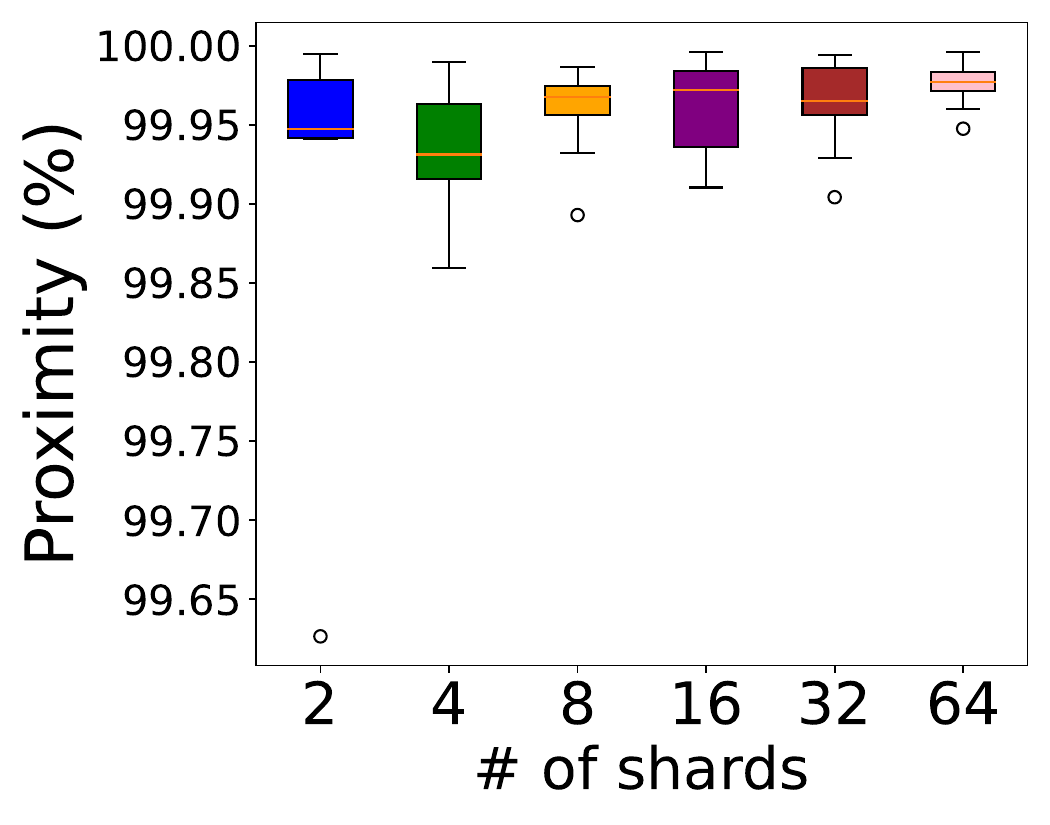}
\label{fig:exp2_distances_to_E}
\end{minipage}
}
\quad
\subfigure[Comparison of the theoretical expectations on TPS and experimental results in the 3 phases of TX handling.]{
\begin{minipage}[t]{0.32\linewidth}
\centering
\includegraphics[width=\textwidth]{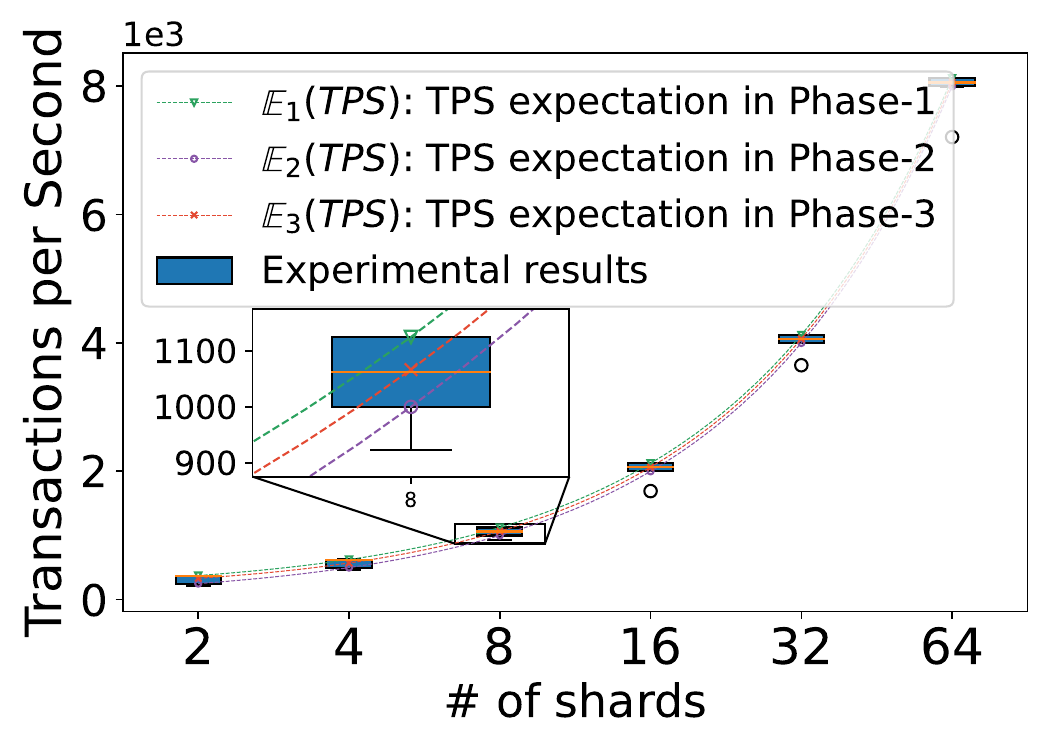} 
\label{fig:exp2-total-box}
\end{minipage}
}
\quad
\subfigure[TCL distributions \hw{\textit{vs}} different shard numbers. The raw TXs are exhausted, and the system shifts to Phase-2 at the \textit{phase boundary}.]{
\begin{minipage}[t]{0.31\linewidth}
\centering
\includegraphics[width=\textwidth]{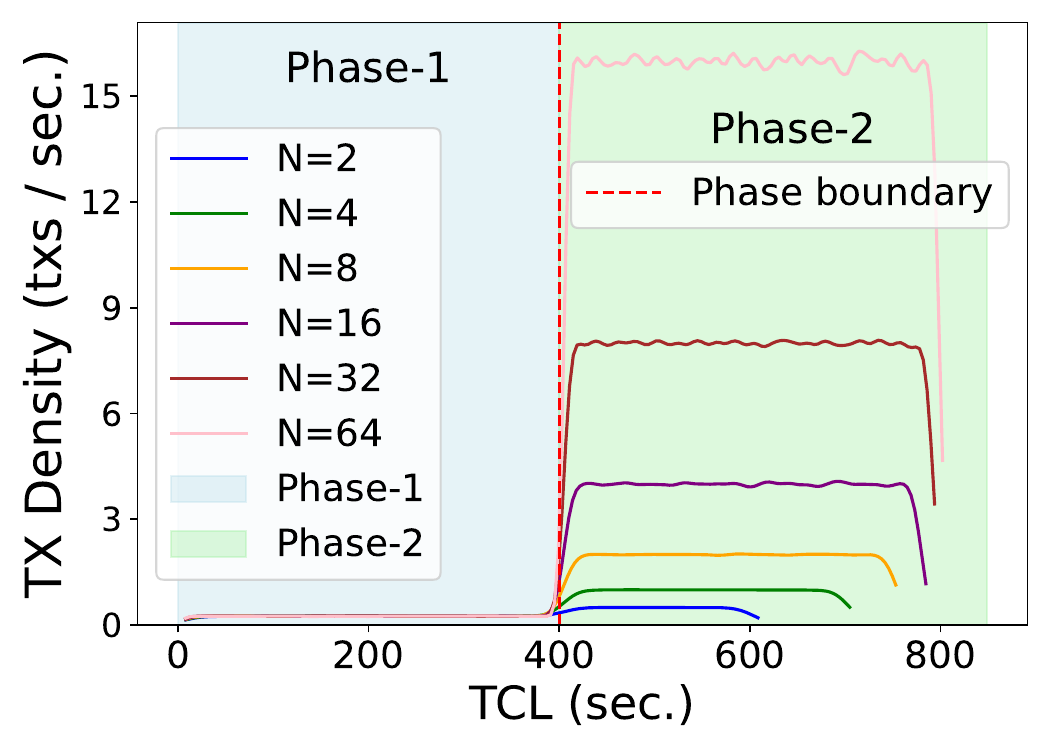}
\label{fig:exp2-TCL-distribution}
\end{minipage}
}
\caption{Correctness verification. Comparison of experimental performance and theoretical derivation with respect to the expected TPS and TCL distribution across different TX-handling phases.}
\end{figure*}

%%%%%%% ---------------------
\subsubsection{Mathematical Analysis}

\textbf{Expected TPS Performance in 3 TX-Handling Phases.} Each shard begins to handle inter-shard relay transactions until all the regular and intra-shard relay transactions are handled in the TX pool. This is because the TX pool is a FIFO queue, and the system collects all transactions before it starts running. Thus, the following 3 phases of TX handling are depicted as follows.

\textit{i)} Phase-1, during which the system processes only raw transactions and adds derived inter-shard relay TXs to the TX pools. \textit{ii)} Phase-2, where only inter-shard relay TXs are processed. \textit{iii)} Phase-3, representing the combined duration of both Phase-1 and Phase-2.
Fig.~\ref {fig: phaseEvolutions} shows the evolution of TX pools in Phase-1 and Phase-2. 
Initially, the system is in Phase-1, and the TX pools comprise only raw TXs, i.e., regular TXs and original CTXs. Once the system has processed and exhausted all raw TXs in the pools, it shifts from Phase-1 to Phase-2. At this stage, the TX pool exclusively contains inter-shard relay TXs.

Tab.~\ref{tab:symbolinExperiments} provides the relevant symbols and explanations. The equation $\mathbb{E}(\text{TPS}) = \mathbb{E}^\prime(\text{TPS}) \times N$ holds due to the balanced workloads of all shards. We can derive from Eq.\eqref{eq2} that $\left | \mathcal{U} \right |=\left | \mathcal{V} \right |=\left | \mathcal{Y} \right |=\frac{N-1}{N} \cdot \left | \mathcal{X} \right |$ , $\left | \mathcal{Z} \right |=\frac{1}{N}\cdot \left | \mathcal{X} \right |$. We can also compute the transaction-packing speed: $\phi=\Theta\cdot N/\delta$.  

Then, let $\mathbb{E}_{1}(\text{TPS})$, $t_1$, $\mathcal{R}_1$ and $\mathcal{W}_1$ denote $\mathbb{E}(\text{TPS})$, $t$, $\mathcal{R}$ and $\mathcal{W}$ in Phase-1, respectively. Similar symbols apply to the other 2 phases. 
The mathematical expectations of TPS of these 3 phases are provided as follows. 
\begin{itemize}
    \item \textbf{Phase-1.} In this phase, the system packs both regular and intra-shard relay transactions into blocks, since inter-shard relay transactions are at the end of the TX pool queues. 
    According to the $\left | \mathcal{V} \right |$ and $\left | \mathcal{Z} \right |$, we can get $\left | \mathcal{R}_1 \right |=\left | \mathcal{V} \right |/2+\left | \mathcal{Z} \right |=\frac{N+1}{2N} \cdot \left | \mathcal{X} \right |$, and $\left | \mathcal{W}_1 \right |=\left | \mathcal{V} \right |+\left | \mathcal{Z} \right |=\left | \mathcal{X} \right |$. 
    The expected time cost $t_1$ is $\left | \mathcal{W}_1 \right |/\phi$. Thus, we have $t_1=\left | \mathcal{W}_1 \right |/\phi=\frac{\delta \cdot \left | \mathcal{X} \right |}{\Theta \cdot N}$. 
    Therefore, the TPS expectation for Phase-1 is expressed as:
    \begin{equation}
            \mathbb{E}_{1}(\text{TPS})=\left | \mathcal{R}_1 \right |/t_1 =\frac{\Theta \cdot (N+1)}{2\delta}.
        \label{eq7}
    \end{equation}
 
    \item \textbf{Phase-2.} In this phase, the system handles the inter-shard relay transactions only. A block packs $\Theta$ transactions, but the credit is only $\Theta/2$. Thus, $\mathbb{E}_{2}(\text{TPS})$ can be calculated by dividing the speed of transaction packing by 2: 
    \begin{equation}
        \begin{aligned}
            \mathbb{E}_{2}(\text{TPS})&=\phi/2=\frac{\Theta \cdot N}{2\delta}. 
        \end{aligned}
        \label{eq8}
    \end{equation}    
    
    \item \textbf{Phase-3.} In the whole TX execution combined phase-1 and phase-2, the number of transactions the system packs is $\left | \mathcal{W}_3 \right | = \left | \mathcal{Z} \right |+\left | \mathcal{U} \right |+\left | \mathcal{V} \right |= \frac{2N-1}{N}\cdot \left | \mathcal{X} \right |$. 
    Here, $t_3$ is the time cost of packing these $\left | \mathcal{W}_3 \right |$ transactions on the chain, and it can be calculated as in Phase-1: $t_3 = \left | \mathcal{W}_3 \right |/\phi= \frac{\delta\cdot (2N-1) \cdot \left | \mathcal{X} \right |}{N^2\cdot \Theta}$. Therefore, the TPS expectation in Phase-3 is written as follows:
    \begin{equation}
        \begin{aligned}
            \mathbb{E}_{3}(\text{TPS}) &= \left | \mathcal{R}_3 \right |/t_3= \left | \mathcal{X} \right | / t_3\\
            &= \frac{N^2\cdot \Theta}{\delta(2N-1)}. 
        \end{aligned}
        \label{eq11}
    \end{equation}  
\end{itemize}

%%%%%%% ---------------------
\textbf{Distributions of TCL for Regular TXs and Cross-shard TXs.}
    Before the system begins processing TXs, random generation makes both regular TXs and cross-shard TXs in the TX pools uniformly distributed. We reuse the previously mentioned Phase-1 and Phase-2 to explain how transactions evolve in the TX pools.
    
    In Phase-1, nodes pack both regular TXs and cross-shard TXs. If a cross-shard TX is packed as an intra-shard relay TX, a corresponding inter-shard relay TX will be sent to another destined shard and appended to the end of that shard's TX pool queue. Therefore, Phase-1 handles only regular TXs. 
    In Phase-2, only inter-shard relay TXs remain in the TX pools. In other words, Phase-2 handles only cross-shard TXs.
    
    Given that the TX pools are FIFO queues and the TX packing speed is a constant $\phi$, the transaction confirmation latency (TCL) for both transaction types is uniformly distributed. In detail, we calculate the TCL distributions for these two types of TXs as follows.

    \begin{itemize}
        \item \textbf{The TCL distribution of regular TXs.} Initially, the mathematical expected number of TXs in all TX pools is $\mathcal{W}$. The expected time duration of Phase-1 is $\frac{\mathcal{W}}{\phi}$. Recall that $\phi=\Theta\cdot N/\delta$, the duration can be written as $\frac{\mathcal{W}\cdot\delta}{N\cdot\Theta}$. As mentioned before, each shard handles regular TXs in Phase-1. Because TXs are randomly generated, the TCL of regular TXs, $\text{TCL}_\mathcal{Z}$, follows a uniform distribution. Thus, we have $\text{TCL}_\mathcal{Z} \sim \text{Uniform}(0, \frac{\mathcal{W}\cdot\delta}{N\cdot\Theta})$.

        \item \textbf{The TCL distribution of cross-shard TXs.} As mathematical expectation, there are a number $\frac{(N-1) \cdot \mathcal{W}}{N}$ of cross-shard TXs being handled in Phase-2. The expected time duration of Phase-2 is $\frac{\mathcal{W} \cdot (N-1)}{\phi \cdot N}$. Therefore, the TCL of cross-shard $\text{TCL}_\mathcal{Y}$ follows a uniform distribution, i.e., $\text{TCL}_\mathcal{Y} \sim \text{Uniform}(\frac{\mathcal{W}\cdot\delta}{N\cdot\Theta}, \frac{\mathcal{W}\cdot\delta}{N\cdot\Theta}+\frac{\mathcal{W} \cdot \delta \cdot(N-1)}{\Theta \cdot N^2})$. 
    \end{itemize}

Since TXs are generated randomly and uniformly, the results indicate that the TCL distributions for both regular TXs and cross-shard TXs follow uniform distributions. If TXs are generated non-uniformly or exhibit bias, the TCL distribution will vary accordingly.

%%%%%%% ---------------------
\subsubsection{Experiment results and comparisons} 

Fig.\ref{fig:exp2-TPS_allShardNumber} shows the piecewise TPS for each epoch when the number of shards varies. The two dashed lines in this figure refer to $\mathbb{E}_{1}(\text{TPS})$ and $\mathbb{E}_{2}(\text{TPS})$, i.e., the TPS expectations for Phase-1 and Phase-2, respectively.
The TPS curves stabilize around the $\mathbb{E}_{1}(\text{TPS})$ and $\mathbb{E}_{2}(\text{TPS})$ lines during Phase-1 and Phase-2, respectively. 
These findings confirm that the experimental results are as expected.

Fig.\ref{fig:exp2_distances_to_E} illustrates the proximity percentages between the observed TPS of each epoch and the expected TPS versus varying shard numbers. The results demonstrate that the observed experimental values closely align with the theoretical expectations for each shard configuration. Fig.~\ref{fig:exp2-total-box} presents the experimental TPS for each epoch alongside the phase-wise TPS expectations. In the magnified subplot, the $\mathbb{E}_{1}(\text{TPS})$ curve aligns with the upper quartile, while the $\mathbb{E}_{2}(\text{TPS})$ curve matches the lower quartile of the box plot. The $\mathbb{E}_{3}(\text{TPS})$ curve crosses between these two.

%%%%%%% --------------------- Fig. 10
\begin{figure}[t]
\centering
\subfigure[TPS and TCL]{
\begin{minipage}[t]{0.52\linewidth}
\centering
\includegraphics[width=\textwidth]{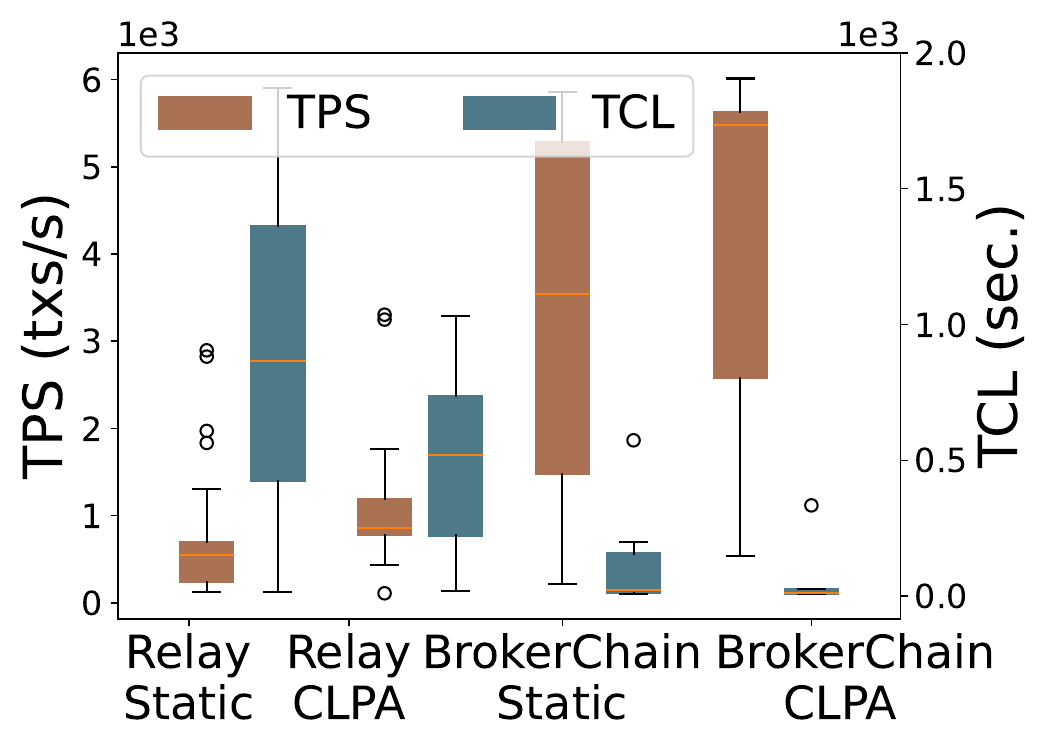}
\label{fig:Exp1_Tcl_Tps_box}
\end{minipage}
}
\subfigure[CTX Ratio]{
\begin{minipage}[t]{0.4\linewidth}
\centering
\includegraphics[width=\textwidth]{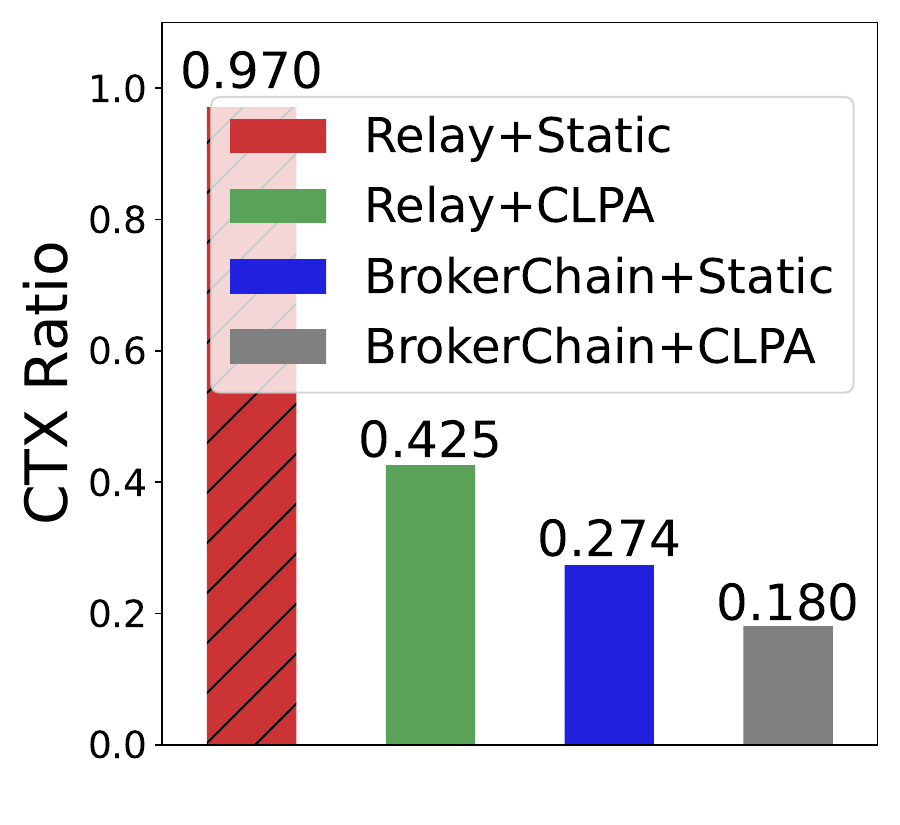} 
\label{fig:Exp1_ctx_ratio}
\end{minipage}
}
\caption{Evaluation on TPS, TCL, and CTX Ratio while invoking 4 mechanisms, i.e., Relay+Static, Relay+CLPA, BrokerChain+Static, and BrokerChain+CLPA.}
\end{figure}
%%%%%%% ---------------------

Fig.~\ref{fig:exp2-TCL-distribution} shows TCL distributions. The total \# of TXs is $\mathcal{W} = 100,000 \times N$. Thus, the vertical red dashed line \textit{phase boundary} is fixed at $\text{TCL} = \frac{\mathcal{W}\cdot\delta}{N\cdot\Theta} = 400^{th}$ second for all shard number configurations. 
To the left of the \textit{phase boundary} line, the system is in Phase-1 and only packages raw TXs. To the right of this line, the system works in Phase-2 and exclusively packages inter-shard relayed TXs.
The observation indicates that TCL follows a uniform distribution in both Phase-1 and Phase-2, which aligns with theoretical analysis.

\begin{figure*}[t]
\centering
\subfigure[The TX pool size of all shards.]{
\begin{minipage}[t]{0.335\linewidth}
\centering
\includegraphics[width=\textwidth]{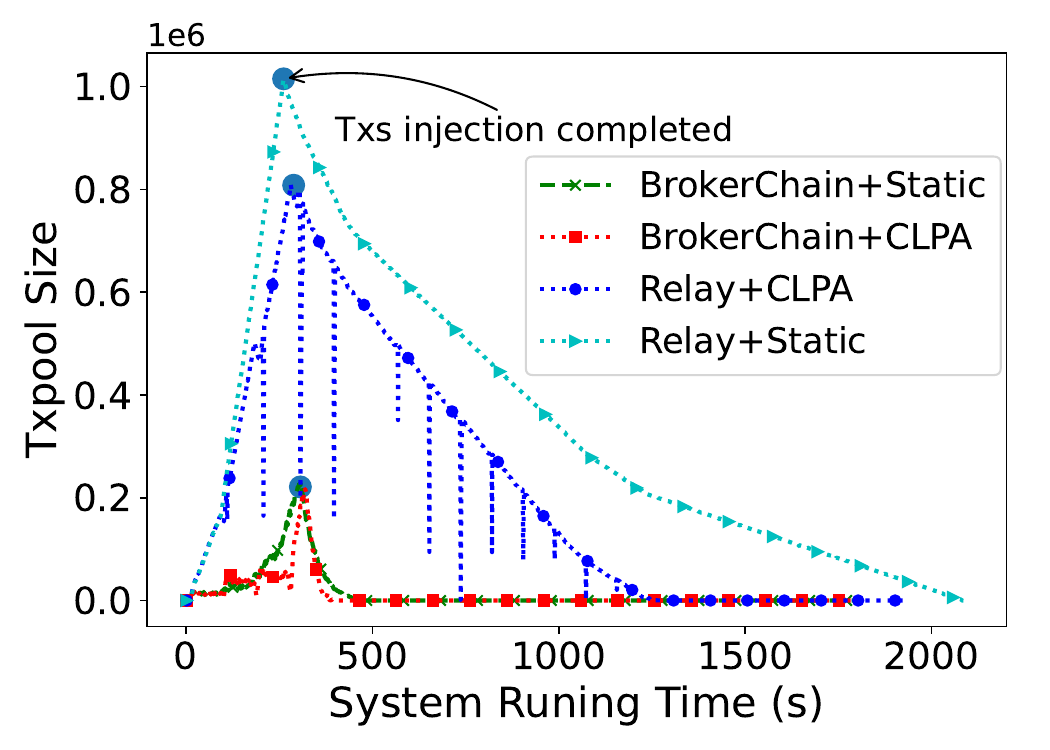}
\label{fig:Exp1_txpool_dynamicInject}
\end{minipage}
}
\quad
\subfigure[Transaction workloads across all shards.]{
\begin{minipage}[t]{0.275\linewidth}
\centering
\includegraphics[width=\textwidth]{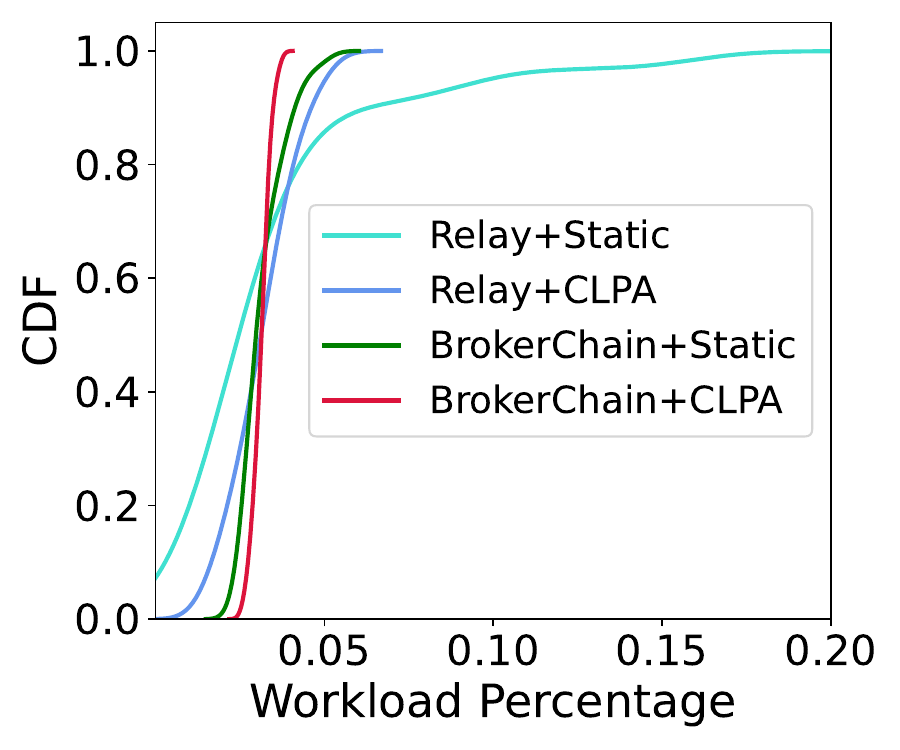} 
\label{fig:Exp1_cdf_workload}
\end{minipage}
}
\quad
\subfigure[\Guang{TPS vs. different transaction injection rates. }]{
\begin{minipage}[t]{0.2875\linewidth}
\centering
\includegraphics[width=\textwidth]{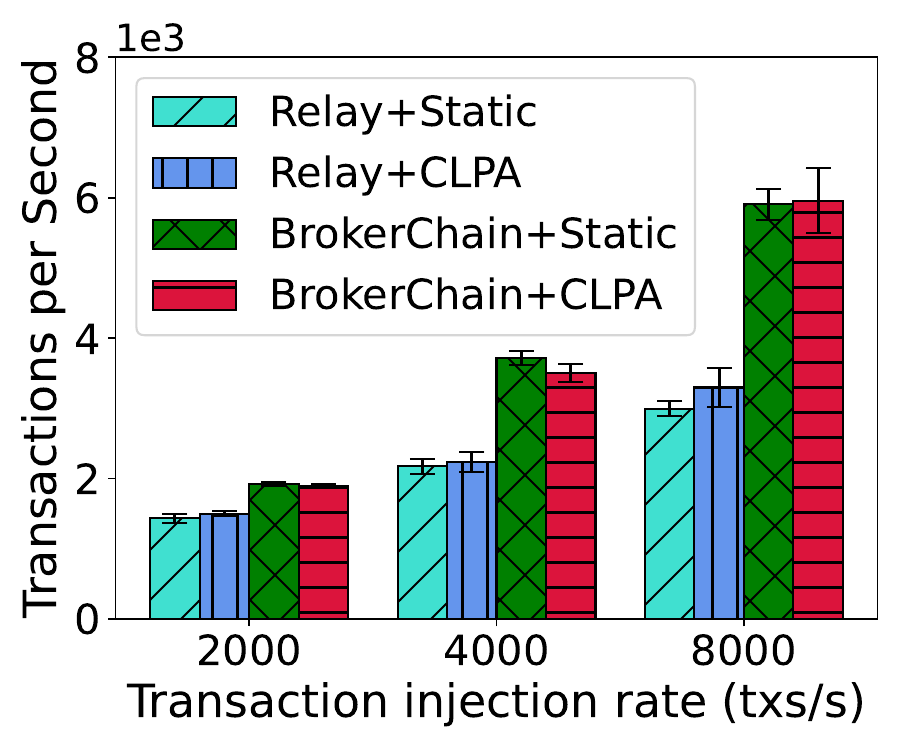} 
\label{fig:Exp3_bar_inject_merge}
\end{minipage}
}
\caption{Evaluation on time-varying TX pool size, workloads, and TPS vs. different TX injection rates.}
\end{figure*}

%%%%%%%%%% ==================
\subsection{Various Metrics Supported to Measure}

To demonstrate \textit{BlockEmulator} 's performance-testing capabilities, we conduct another set of experiments using approximately 1.6 million historical Ethereum transactions~\cite{Zheng2020xblock} as the dataset. The number of shards is set to 32, and the number of nodes in each shard is set to 4. In other words, we run 128 nodes in our physical machines. In addition, we fix the transaction injection rate at 4,000 txs/s by default and increase it by 4,000 txs/s each epoch.

To highlight the features that support secondary development, we also implement some state-of-the-art mechanisms in our \textit{BlockEmulator} and compare them using different metrics. We introduce the implemented mechanisms as follows. 
\begin{itemize}
    \item \textbf{Relay+Static.} The system handles cross-shard transactions with \textit{transaction relay mechanism}\cite{wang2019monoxide}. The account states are divided into different shards by a static sharding method.  
    
    \item \textbf{Relay+CLPA.} The cross-shard transaction handling method is \textit{transaction relay mechanism}. \Guang{The account states are redistributed using CLPA\cite{li2022achieving}, which is a dynamic sharding algorithm.} 
    
    \item \textbf{BrokerChain+Static.} BrokerChain protocol \cite{huang2022brokerchain} is adopted to handle cross-shard transactions. The account states are sharded statically. 
    
    \item \textbf{BrokerChain+CLPA.} This mechanism integrates the \hhw{solutions} of broker accounts and CLPA. 
\end{itemize}

In our experiments, we selected the top 10 accounts that transfer the most transactions as broker accounts. These broker accounts are involved in 80\% of total transactions. 
With these well-selected broker accounts, the BrokerChain-related mechanisms thus perform better.
We use an offline version of the CLPA algorithm on partial data and find that the CTX Ratio is reduced by about 55\%. 
Therefore, we expect that \textit{BrokerChain+Static} will outperform \textit{Relay+CLPA}, and \textit{BrokerChain+CLPA} will show better performance than \textit{BrokerChain+Static}.

As shown in Fig.~\ref{fig:Exp1_Tcl_Tps_box}, \textit{BrokerChain+Static} outperforms \textit{Relay+CLPA}, and both perform better than \textit{Relay+Static} in terms of TPS and TCL. The results of this group of experiments perfectly match what we predicted in the previous paragraph.

Interestingly, \textit{BrokerChain+CLPA} does not show a dominating advantage over \textit{BrokerChain+Static}.
One possible reason is that the \textit{CLPA} algorithm requires a reconfiguration stage for account migration, which adds time overhead. 
Another reason can be inferred from Fig.~\ref{fig:Exp1_ctx_ratio}, which implies that \textit{Relay+CLPA} has a much lower CTX ratio than \textit{Relay+Static}. However, \textit{BrokerChain+CLPA} has a slightly lower CTX ratio than \textit{BrokerChain+Static}. This means the \textit{BrokerChain} mechanism has already significantly reduced the CTX Ratio, so the impact of running \textit{CLPA} is not apparent.

Fig.~\ref{fig:Exp1_txpool_dynamicInject} shows that the cumulative size of TX pools varies over time. Note that, in \textit{BrokerChain+CLPA} and \textit{Relay+CLPA}, TX pool size suddenly drops and rises at certain time slots. This is because of the transaction mitigation operations in \textit{CLPA}. In addition, the TX pool size performance of both \textit{BrokerChain+Static} and \textit{BrokerChain+CLPA} increases slowly but declines fast. This suggests that these two mechanisms yield higher TPS and fewer accumulated transactions in TX pools than the other two mechanisms.

To further explore why \textit{BrokerChain+CLPA} and \textit{BrokerChain+Static} perform better in Fig.~\ref{fig:Exp1_txpool_dynamicInject}, we evaluate the workload distribution across shards. Fig.~\ref{fig:Exp1_cdf_workload} shows the results. 
We calculate each shard's workload by summing the packed transactions in all committed blocks within the shard.  
\textit{BrokerChain+CLPA} shows the most balanced workloads because its CDF curve rises rapidly. This indicates a high density of workload percentages, suggesting balanced workloads across shards. 
The CDF curve of \textit{Relay+Static} grows slowly, while the CDFs of the other 3 mechanisms reach $1.0$ fast. This shows that both \textit{CLPA} and \textit{BrokerChain} balance workloads effectively.

From Fig.~\ref{fig:Exp1_txpool_dynamicInject}, we find that the curves of \textit{Relay+CLPA} and \textit{Relay+Static} are close for the first 200 seconds.
The curves of \textit{BrokerChain+CLPA} and \textit{BrokerChain+Static} are almost the same. 
These observations are interesting. Thus, we conducted another experiment to find some clues. This group of experiments runs for 1000 seconds. We record TPS every epoch (i.e., 80 seconds). To get adequate experimental data, we use about 8 million historical transactions collected from Ethereum~\cite{Zheng2020xblock}.

Fig.~\ref{fig:Exp3_bar_inject_merge} demonstrates the TPS of the 4 mechanisms at different transaction injection rates.
Surprisingly, \textit{CLPA} has little or a negative effect when integrated with the \textit{Relay} and \textit{Brokerchain} mechanisms. This is because \textit{CLPA} improves TPS by moving transactions from the overloaded shards to the underloaded ones. However, when the transaction injection rate is low, there are almost no overloaded shards. Similarly, there is almost no underloaded shard when the transaction injection rate is high. In addition, the reconfiguration time consumed by \textit{CLPA} also degrades performance. Therefore, \textit{CLPA} performs better if the transaction injection rate is low and TX pools are full. 
In Fig.~\ref{fig:Exp1_txpool_dynamicInject}, \textit{CLPA} outperforms when transaction injection stops. This observation confirms our aforementioned explanation.

Finally. when the transaction injection rates are relatively low, the TPS of BrokerChain+CLPA and BrokerChain+Static are very close to the injection rates.
This is because \textit{BrokerChain} can shift a cross-shard transaction to a regular transaction if its payer or payee is a broker account. The proper broker accounts we selected for experiments help these two BrokerChain-related mechanisms work well.

%%%%%%%%%% ==================
%%%%%%%%%% ==================
%%%%%%%%%% ==================
\section{In-depth Discussion}

In this section, we discuss the extra functionalities and benefits of BlockEmulator in depth.

\subsection{Extra Functionalities of BlockEmulator}

 \textbf{State Machine Replication.} BlockEmulator achieves complete state machine replication at each node while implementing the PBFT protocol. During operation, it outputs the Merkle root of the state trie. By comparing the Merkle roots of the state tries across different nodes within the same shard, users can verify the consistency of these Merkle roots. Each node runs the same PBFT consensus protocol, making this comparison meaningful. This consistency confirms that the platform successfully realizes complete state machine replication across all nodes within a shard. Furthermore, nodes classify a block as invalid if the Merkle root in that block does not match the local Merkle root.

 \textbf{View-change in PBFT.} View-change operation is a secure mechanism in PBFT. To implement this, we set a timer for each node. When a block is committed, the timer resets and starts counting down again. If the nodes within a shard fail to reach consensus, the timer will count to zero. At that point, these nodes initiate a view-change proposal to change the PBFT leader, ensuring blockchain liveness. Users can adjust the timeout threshold that activates the PBFT view-change mechanism by configuring the \texttt{PbftViewChangeTimeOut} parameter. The implementation code for the view-change mechanism is available in our open-source GitHub project\footnote{\url{https://github.com/HuangLab-SYSU/block-emulator/blob/main/consensus\_shard/pbft\_all/view\_change.go}}.

\subsection{Benefits and Features of BlockEmulator}

% \subsubsection{Verisimilitude of BlockEmualtor}

 \textbf{Verisimilitude.} BlockEmulator enables the simulation of network characteristics such as latency and bandwidth constraints. Users can further refine network performance by leveraging operating system tools to specify node behaviors.
 BlockEmulator can also replay real Ethereum transactions. Users can validate the correctness of simulations using the verification scripts available in the \texttt{./query} directory of its GitHub repository.
    In addition, BlockEmulator emulates the primary roles in blockchain systems, including consensus nodes (i.e., worker nodes), transaction initiators (embedded within the \textit{supervisor} node), and committees (also embedded within the \textit{supervisor} node).
    The evaluation results in Section \ref{sec:Experiments} show that BlockEmulator's results closely align with theoretical analysis. For example, Fig.~\ref{fig:exp2_distances_to_E} shows that, with 64 shards, BlockEmulator achieves around 99.9\% proximity between theoretical analysis and experimental TPS.

% \subsubsection{Extendibility of BlockEmulator}

\textbf{Extendibility.}
BlockEmulator uses a layered architecture, enabling users to customize specific layers based on their research goals. BlockEmulator offers many interfaces to support \textit{secondary development} and extend functionality.
Several academic articles \cite{chen2024broker2earn, huang2024account, zheng2025justitia, ding2024presto, li2024spring, song2025aero} have leveraged BlockEmulator to implement and evaluate blockchain algorithms, protocols, and mechanisms. This demonstrates BlockEmulator's adaptability and extensibility.

% \subsubsection{Scalability of BlockEmulator} 

\textbf{Scalability.}
% 这部分的回复思路是这样的：
    % 1. BlockEmulator 的可扩展性主要取决于 “共识协议”，因为共识协议能够决定共识节点间的通信复杂度，进而决定了可扩展性。BlockEmulator 支持用户自定义的共识机制，这允许研究者们探索、实现更好的共识机制。
    %
    The scalability of BlockEmulator primarily depends on consensus protocols, which determine the communication complexity among consensus nodes. BlockEmulator supports user-defined consensus, allowing researchers to implement efficient intra-shard and cross-shard protocols to emulate a large blockchain network.

    % 2. （虽然可扩展性主要取决于 “共识协议”，但我们在设计 BlockEmulator 时，也有考虑大规模部署时会遇到的问题）在网络层中，BlockEmulator 采用了 TCP 长连接（keep-alive connections）。这可以减少共识节点间重复建立和关闭 TCP 连接的开销，进而提高了可扩展性。在控制层中，BlockEmulator 将不同功能集成在 supervisor 节点中，在大规模实验里 supervisor 可能由于集成了太多功能而成为性能瓶颈。但由于我们在集成功能时采用了模块化设计，所以用户可以轻松地按照 supervisor 的不同功能，将 supervisor 拆分为若干个具备不同功能的节点，从而减轻 supervisor 的单点负载。

    Although scalability is primarily associated with consensus protocols, we also consider how to support large-scale deployments when designing BlockEmulator. In the network layer, BlockEmulator utilizes keep-alive TCP connections. This approach reduces the overhead of repeatedly reestablishing  TCP connections between consensus nodes, thereby improving scalability.
    In the control layer, BlockEmulator integrates various functionalities into the supervisor node. If too many functions are integrated, the supervisor may become a bottleneck in very large-scale experiments. 
    We considered this and adopted a modular design when integrating these functions. This lets users easily split the supervisor into multiple units, depending on their design. This split can effectively alleviate the single-node bottleneck.

\begin{figure}[h!t]
    \centering
    \includegraphics[width=0.9\linewidth]{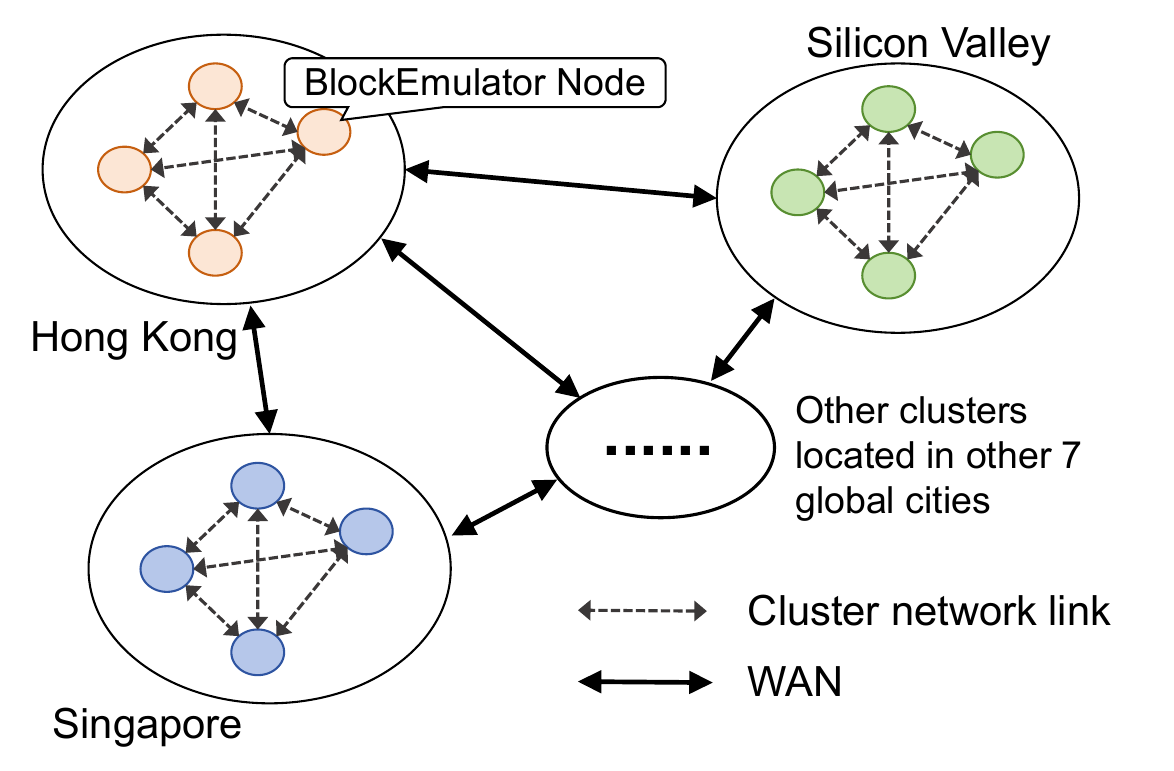}
    \caption{A real use case of deploying BlockEmulator nodes across globally deployed clusters. These clusters are located in the following 10 cities: Hong Kong in China, Seoul in South Korea, Tokyo in Japan, Singapore, Bangkok in Thailand, Mumbai in India, Silicon Valley in the United States, Frankfurt in Germany, Virginia in the United States, and São Paulo in Brazil.} 
    \label{fig:networkExplain}
\end{figure}

    % 3. 举例说明 BlockEmulator 可以支持大规模的部署，用例子说明 BlockEmulator 具备可扩展性。
    Fig.~\ref{fig:networkExplain} shows a real experimental case of deploying BlockEmulator nodes across 10 different clusters. Specifically, we deployed BlockEmulator on Tencent Cloud using 100 virtual machines (VMs). Each VM instance hosts one consensus node featuring 4 cores of Intel Xeon 2.5GHz CPU. The WAN network bandwidth is set to 100 Mbps. We launched 100 VM instances across 10 major cities worldwide, with 10 instances in each cluster.

    In this real use case, BlockEmualtor users can control delay and bandwidth by modifying network parameters. Globally deployed clusters connect via WAN, where real-world network conditions significantly affect blockchain node performance. Depending on the network characteristics, users can deploy nodes in various environments and configure network parameters to meet their specific needs. The deployment scale depends on the experiment budget using BlockEmulator. If the budget is sufficient, users can deploy many BlockEmulator instances in cloud environments worldwide.

% \subsubsection{Supporting Smart Contracts}

 \textbf{Supporting Smart Contracts.} The current version of BlockEmulator does not support evaluating smart contracts in a blockchain-sharding environment because this is a challenging research topic. We hope to add this feature in a future version of BlockEmulator.

%%%%%%%%%% ==================
%%%%%%%%%% ==================
%%%%%%%%%% ==================
\section{Conclusion}\label{sec:conclusion}

BlockEmulator is designed as a testbed for blockchain researchers. It supports basic operations in existing blockchains and offers powerful programming interfaces for evaluating new blockchain protocols or consensus algorithms. In particular, it enables researchers to develop and evaluate new customized blockchain-sharding protocols. To verify the correctness of the experiment data yielded by BlockEmulator, we compared the observed performance with theoretical calculations.
Other experiments show that BlockEmulator supports meaningful performance testing using typical metrics in a blockchain sharding system, including TPS, transaction confirmation latency, cross-shard transaction ratio, and transaction-pool queue length.

We have open-sourced BlockEmulator on Github~\cite{blockemulatorGitHub}. This project has gained over 350 stars. We hope BlockEmulator becomes a practical and useful experimental tool for blockchain researchers who need to implement and test new protocols, algorithms, and mechanisms.
In future versions of BlockEmulator, we plan to support smart-contract functionality.

%%%%%%%%%% ==================
%%%%%%%%%% ==================
%%%%%%%%%% ==================
\section*{Acknowledgements}
The authors appreciate all the great efforts and contributions to BlockEmulator made by the students who have graduated from Prof. Huang's Research Group (a.k.a. HuangLab), including Xiaowen~Peng, Jianzhou~Zhan, Shenyang~Zhang, Canlin~Li, Yue~Lin, Yetong~Zhao, Miaoyong~Xu, Guang~Ye, Zhaokang~Yin, and Qinde~Chen, as well as other current team members who are not shown in the author list, such as Jianwen~Yuan, Jianbo~Xiong, Yang~Zhou, Junhao~Wu, Feihong~Hu, and Baozhou~Xie.

\bibliographystyle{IEEEtran}
\bibliography{references}

@inproceedings{huang2022brokerchain,
  title={Brokerchain: A cross-shard blockchain protocol for account/balance-based state sharding},
  author={Huang, Huawei and Peng, Xiaowen and Zhan, Jianzhou and Zhang, Shenyang and Lin, Yue and Zheng, Zibin and Guo, Song},
  booktitle={Proc. of IEEE Conference on Computer Communications (INFOCOM)},
  pages={1968--1977},
  year={2022},
  organization={IEEE}
}

@inproceedings{wang2019monoxide,
  title={Monoxide: Scale out Blockchains with Asynchronous Consensus Zones.},
  author={Wang, Jiaping and Wang, Hao},
  booktitle={Proc. of 20th USENIX Symposium on Networked Systems Design and Implementation},
  volume={2019},
  pages={95--112},
  year={2019}
}

@inproceedings{dinh2017blockbench,
  title={Blockbench: A framework for analyzing private blockchains},
  author={Dinh, Tien Tuan Anh and Wang, Ji and Chen, Gang and Liu, Rui and Ooi, Beng Chin and Tan, Kian-Lee},
  booktitle={Proc. of the 2017 ACM international conference on management of data},
  pages={1085--1100},
  year={2017}
}

@inproceedings{aoki2019simblock,
  title={Simblock: A blockchain network simulator},
  author={Aoki, Yusuke and Otsuki, Kai and Kaneko, Takeshi and Banno, Ryohei and Shudo, Kazuyuki},
  booktitle={Proc. of IEEE Conference on Computer Communications Workshops (INFOCOM WKSHPS)},
  pages={325--329},
  year={2019},
  organization={IEEE}
}

@inproceedings{stoykov2017vibes,
  title={Vibes: fast blockchain simulations for large-scale peer-to-peer networks},
  author={Stoykov, Lyubomir and Zhang, Kaiwen and Jacobsen, Hans-Arno},
  booktitle={Proc. of the 18th ACM/IFIP/USENIX Middleware Conference: Posters and Demos},
  pages={19--20},
  year={2017}
}

@inproceedings{qi2022sstore,
  title={{S-Store: A Scalable Data Store towards Permissioned Blockchain Sharding}},
  author={Qi, Xiaodong},
  booktitle={Proc. of IEEE Conference on Computer Communications (INFOCOM)},
  pages={1978--1987},
  year={2022},
  organization={IEEE}
}

@inproceedings{yang2022dispersedledger,
  title={{DispersedLedger: High-Throughput Byzantine Consensus on Variable Bandwidth Networks}},
  author={Yang, Lei and Park, Seo Jin and Alizadeh, Mohammad and Kannan, Sreeram and Tse, David},
  booktitle={Proc. of 19th USENIX Symposium on Networked Systems Design and Implementation},
  pages={493--512},
  year={2022}
}

@inproceedings{li2022achieving,
  title={{Achieving Scalability and Load Balance across Blockchain Shards for State Sharding}},
  author={Li, Canlin and Huang, Huawei and Zhao, Yetong and Peng, Xiaowen and Yang, Ruijie and Zheng, Zibin and Guo, Song},
  booktitle={Proc. of 2022 41st International Symposium on Reliable Distributed Systems (SRDS)},
  pages={284--294},
  year={2022},
  organization={IEEE}
}

@article{king2012ppcoin,
  title={Ppcoin: Peer-to-peer crypto-currency with proof-of-stake},
  author={King, Sunny and Nadal, Scott},
  journal={self-published paper, August},
  volume={19},
  number={1},
  year={2012}
}

@inproceedings{zhang2023txallo,
  title={TxAllo: Dynamic Transaction Allocation in Sharded Blockchain Systems},
  author={Zhang, Yuanzhe and Pan, Shirui and Yu, Jiangshan},
  booktitle={Proc. of IEEE 39th International Conference on Data Engineering (ICDE)},
  pages={721--733},
  year={2023},
  organization={IEEE}
}

@article{zhang2019double,
  title={Double-spending with a sybil attack in the bitcoin decentralized network},
  author={Zhang, Shijie and Lee, Jong-Hyouk},
  journal={IEEE transactions on Industrial Informatics},
  volume={15},
  number={10},
  pages={5715--5722},
  year={2019},
  publisher={IEEE}
}

@inproceedings{gao2019power,
  title={Power adjusting and bribery racing: Novel mining attacks in the bitcoin system},
  author={Gao, Shang and Li, Zecheng and Peng, Zhe and Xiao, Bin},
  booktitle={Proc. of the 2019 ACM SIGSAC Conference on Computer and Communications Security},
  pages={833--850},
  year={2019}
}

@article{trybulec1990pigeon,
  title={Pigeon hole principle},
  author={Trybulec, Wojciech A},
  journal={Journal of Formalized Mathematics},
  volume={2},
  number={199},
  pages={0},
  year={1990}
}

@inproceedings{luu2016secure,
  title={A secure sharding protocol for open blockchains},
  author={Luu, Loi and Narayanan, Viswesh and Zheng, Chaodong and Baweja, Kunal and Gilbert, Seth and Saxena, Prateek},
  booktitle={Proc. of the 2016 ACM SIGSAC conference on computer and communications security},
  pages={17--30},
  year={2016}
}

@inproceedings{zamani2018rapidchain,
  title={{Rapidchain: Scaling blockchain via full sharding}},
  author={Zamani, Mahdi and Movahedi, Mahnush and Raykova, Mariana},
  booktitle={Proc. of the 2018 ACM SIGSAC conference on computer and communications security},
  pages={931--948},
  year={2018}
}

@article{brown2018corda,
  title={The corda platform: An introduction},
  author={Brown, Richard Gendal},
  journal={Retrieved},
  volume={27},
  pages={2018},
  year={2018}
}

@article{mohanty2018deploying,
  title={Deploying smart contracts},
  author={Mohanty, Debajani and Mohanty, Debajani},
  journal={Ethereum for Architects and Developers: With Case Studies and Code Samples in Solidity},
  pages={105--138},
  year={2018},
  publisher={Springer}
}

@inproceedings{androulaki2018hyperledger,
  title={{Hyperledger fabric: a distributed operating system for permissioned blockchains}},
  author={Androulaki, Elli and Barger, Artem and Bortnikov, Vita and Cachin, Christian and Christidis, Konstantinos and De Caro, Angelo and Enyeart, David and Ferris, Christopher and Laventman, Gennady and Manevich, Yacov and others},
  booktitle={Proc. of the thirteenth EuroSys conference (EuroSys)},
  pages={1--15},
  year={2018}
}

@misc{Open2023Fastest,
author={OpenEthereum},
title={{The Fastest and most Advanced Ethereum Client.}},
year={Retrieved: May 23, 2023},
howpublished={\url{https://github.com/openethereum/parity-ethereum}}
}

@book{prusty2017building,
  title={Building blockchain projects},
  author={Prusty, Narayan},
  year={2017},
  publisher={Packt Publishing Ltd}
}

@article{Zheng2020xblock,
author = {Zheng, Peilin and Zheng, Zibin and Wu, Jiajing and Dai, Hong-Ning},
journal = {IEEE Open J. Comput. Soc.},
pages = {95--106},
publisher = {IEEE},
title = {{XBlock-ETH: Extracting and exploring blockchain data from Ethereum}},
volume = {1},
month = {May},
year = {2020}
}

@inproceedings{li2023cochain,
  title={{CoChain: High Concurrency Blockchain Sharding via Consensus on Consensus}},
  author={Li, Mingzhe and Lin, You and Zhang, Jin and Wang, Wei},
  booktitle={Proc. of IEEE Conference on Computer Communications (INFOCOM)},
  pages={1--10},
  year={2023},
}

@article{li2023lbchain,
  title={{LB-Chain: Load-Balanced and Low-Latency Blockchain Sharding via Account Migration}},
  author={Li, Mingzhe and Wang, Wei and Zhang, Jin},
  journal={IEEE Transactions on Parallel and Distributed Systems},
  year={2023},
  publisher={IEEE}
}

@article{hong2023gridb,
  title={{GriDB: Scaling Blockchain Database via Sharding and Off-Chain Cross-Shard Mechanism}},
  author={Hong, Zicong and Guo, Song and Zhou, Enyuan and Chen, Wuhui and Huang, Huawei and Zomaya, Albert},
  journal={Proceedings of the VLDB Endowment},
  volume={16},
  number={7},
  pages={1685--1698},
  year={2023},
  publisher={VLDB Endowment}
}

@misc{blockemulatorGitHub,
author    = { },
title     = {{BlockEmulator}},
url       = {https://github.com/HuangLab-SYSU/block-emulator},
}

@inproceedings{dwork2005pebbling,
  title={Pebbling and proofs of work},
  author={Dwork, Cynthia and Naor, Moni and Wee, Hoeteck},
  booktitle={Advances in Cryptology--CRYPTO 2005: 25th Annual International Cryptology Conference, Santa Barbara, California, USA, August 14-18, 2005. Proceedings 25},
  pages={37--54},
  year={2005},
  organization={Springer}
}

@article{Wood2014Ethereum,
  title={Ethereum: A secure decentralised generalised transaction ledger},
  author={Wood, Gavin and others},
  journal={Ethereum project yellow paper},
  volume={151},
  number={2014},
  pages={1--32},
  year={2014}
}

@article{al2017chainspace,
  title={Chainspace: A sharded smart contracts platform},
  author={Al-Bassam, Mustafa and Sonnino, Alberto and Bano, Shehar and Hrycyszyn, Dave and Danezis, George},
  journal={arXiv preprint arXiv:1708.03778},
  year={2017}
}

@inproceedings{kokoris2018omniledger,
  title={Omniledger: A secure, scale-out, decentralized ledger via sharding},
  author={Kokoris-Kogias, Eleftherios and Jovanovic, Philipp and Gasser, Linus and Gailly, Nicolas and Syta, Ewa and Ford, Bryan},
  booktitle={Proc. of IEEE symposium on security and privacy (S\&P)},
  pages={583--598},
  year={2018},
}

@inproceedings{dang2019towards,
  title={Towards scaling blockchain systems via sharding},
  author={Dang, Hung and Dinh, Tien Tuan Anh and Loghin, Dumitrel and Chang, Ee-Chien and Lin, Qian and Ooi, Beng Chin},
  booktitle={Proc. of the 2019 international conference on management of data},
  pages={123--140},
  year={2019}
}

@inproceedings{huang2024account,
  title={Account Migration across Blockchain Shards using Fine-tuned Lock Mechanism},
  author={Huang, Huawei and Lin, Yue and Zheng, Zibin},
  booktitle={IEEE Conference on Computer Communications (INFOCOM)},
  pages={271--280},
  year={2024},
  organization={IEEE}
}

@inproceedings{chen2024broker2earn,
  title={Broker2Earn: towards maximizing broker revenue and system liquidity for sharded blockchains},
  author={Chen, Qinde and Huang, Huawei and Yin, Zhaokang and Ye, Guang and Yang, Qinglin},
  booktitle={Proc. of IEEE Conference on Computer Communications (INFOCOM)},
  pages={251--260},
  year={2024},
  organization={IEEE}
}

@inproceedings{zheng2025justitia,
  title={{Justitia: An Incentive Mechanism towards the Fairness of Cross-shard Transactions}},
  author={Zheng, Jian and Huang, Huawei and Liu, Yinqiu and Li, Taotao and Dai, Hong-Ning and Zheng, Zibin},
  booktitle={Proc. of IEEE Conference on Computer Communications (INFOCOM)},
  pages={1--10},
  year={2025},
}

@inproceedings{ding2024presto,
  title={Presto: Optimizing Cross-Shard Transactions in Sharded Blockchain Architecture},
  author={Ding, Qiuyu and Zhang, Rongkai and Yin, Shenglin and Li, PengZe and Guan, Shengjie and Xiao, Zhen and Long, Jieyi},
  booktitle={Proc. of 43rd International Symposium on Reliable Distributed Systems (SRDS)},
  pages={139--149},
  year={2024},
  organization={IEEE}
}

@inproceedings{li2024spring,
  title={SPRING: Improving the Throughput of Sharding Blockchain via Deep Reinforcement Learning Based State Placement},
  author={Li, Pengze and Song, Mingxuan and Xing, Mingzhe and Xiao, Zhen and Ding, Qiuyu and Guan, Shengjie and Long, Jieyi},
  booktitle={Proc. of the ACM on Web Conference 2024 (WWW'24)},
  pages={2836--2846},
  year={2024}
}

@inproceedings{song2025aero,
  title={AERO: Enhancing Sharding Blockchain via Deep Reinforcement Learning for Account Migration},
  author={Song, Mingxuan and Li, Pengze and Zhou, Bohan and Yin, Shenglin and Xiao, Zhen and Long, Jieyi},
  booktitle={Proc. of the ACM on Web Conference 2025 (WWW'25)},
  pages={1--11},
  year={2025}
}

% biography section

% %%%% ====================

\begin{IEEEbiography}[{\includegraphics[width=1in,height=1.25in,clip,keepaspectratio]{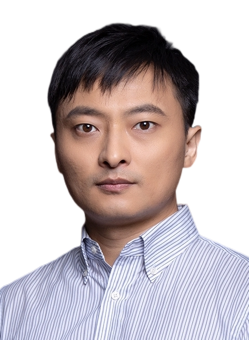}}]{Huawei~Huang} (Senior Member, IEEE) received his Ph.D. in computer science and engineering from the University of Aizu, Japan. He is a professor in the School of Software Engineering, Sun Yat-sen University, China. His research interests mainly include blockchain systems and protocols. He has served as a guest editor for multiple blockchain special issues in the IEEE Journal on Selected Areas in Communications, IEEE Open Journal of the Computer Society, and IET Blockchain. He also served as TPC chair for blockchain conferences, such as IEEE Global Blockchain Conference, IEEE Services, ICWS, BlockSys, etc. He has been recognized on the yearly World's Top 2\% Scientists list since 2021.
\end{IEEEbiography}

\begin{IEEEbiography}[{\includegraphics[width=1in,height=1.25in,clip,keepaspectratio]{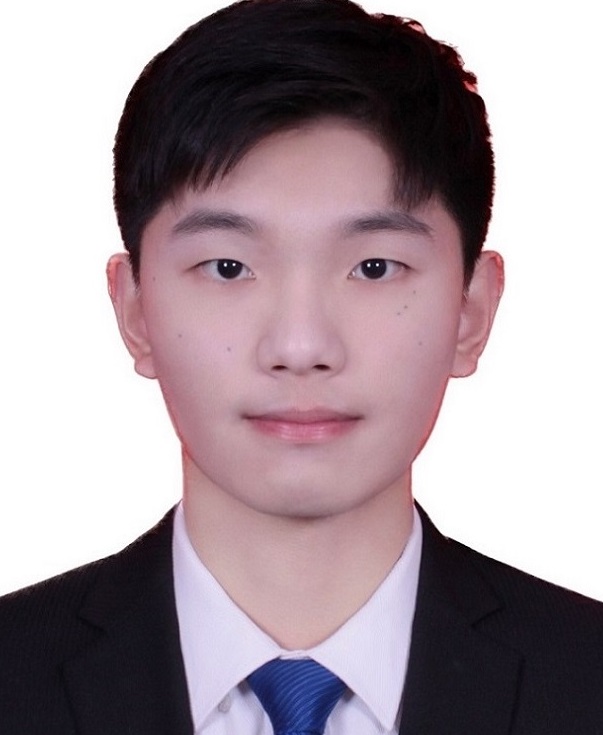}}]
{Guang~Ye} received a master's degree in Software Engineering from Sun Yat-sen University. His research interests mainly include Blockchain.
\end{IEEEbiography}

\begin{IEEEbiography}[{\includegraphics[width=1in,height=1.25in, clip,keepaspectratio]{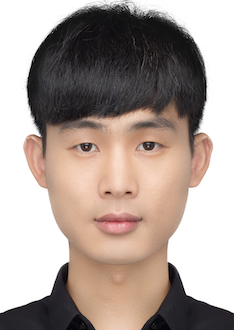}}]{Qinglin~Yang} received a Ph.D. degree in computer science and engineering from the University of Aizu, Japan, in 2021. He has served as a guest editor for a blockchain special issue in the IEEE Open Journal of the Computer Society. He also served as a TPC chair for several conferences, including blockchain and machine learning. He has worked as a research fellow at Sun Yat-sen University, China. He is an assistant professor with the Cyberspace Institute of Advanced Technology, Guangzhou University, China. His current research interests include federated learning, Web3, and blockchain.
\end{IEEEbiography}

\begin{IEEEbiography}[{\includegraphics[width=1in,height=1.25in,clip,keepaspectratio]{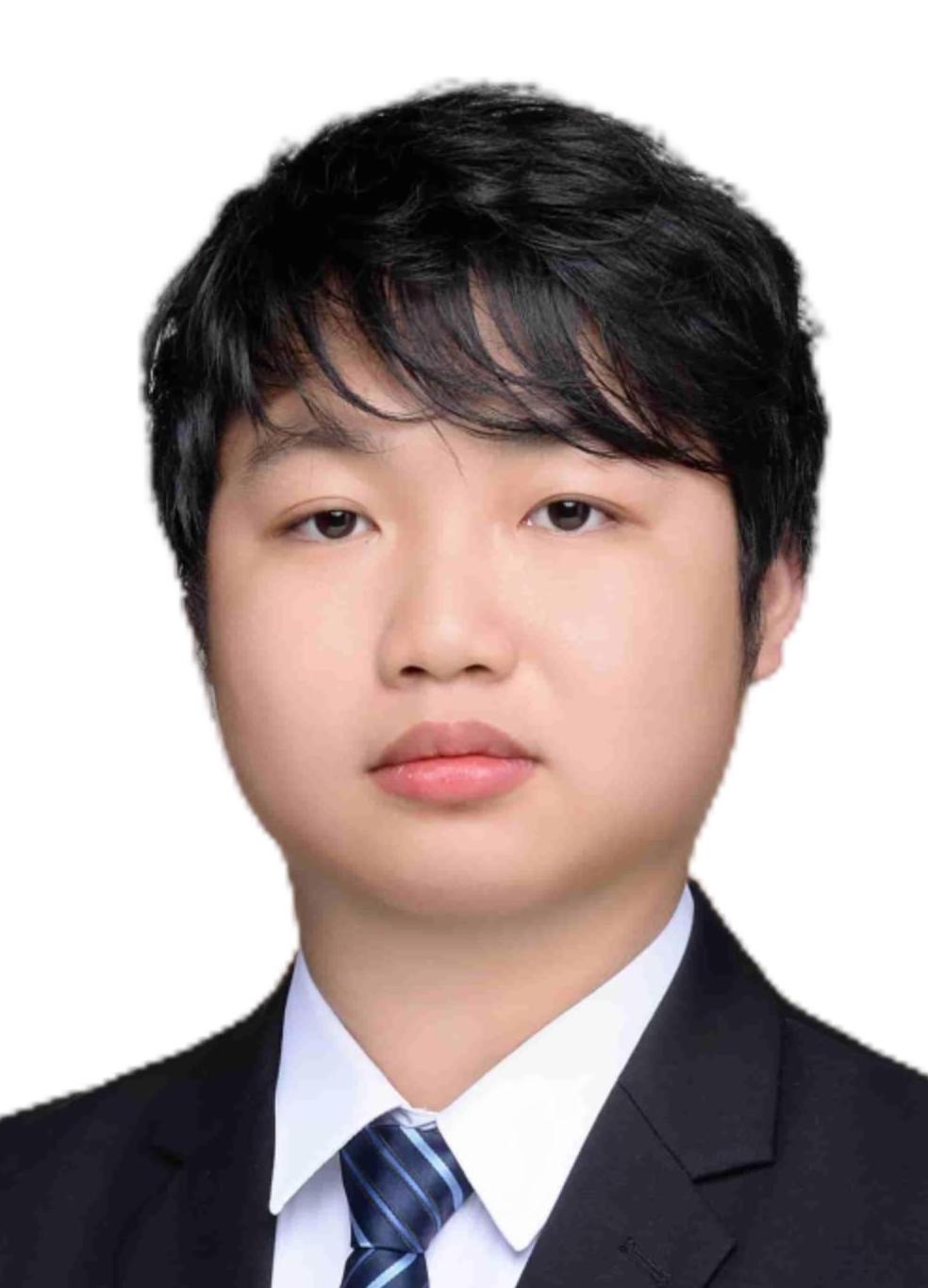}}]
{Qinde~Chen} received a Ph.D. degree from the School of Software Engineering, Sun Yat-sen University. He is now a postdoctoral researcher at HKUST. His research interests mainly focus on blockchain.
\end{IEEEbiography}

\begin{IEEEbiography}[{\includegraphics[width=1in,height=1.25in,clip,keepaspectratio]{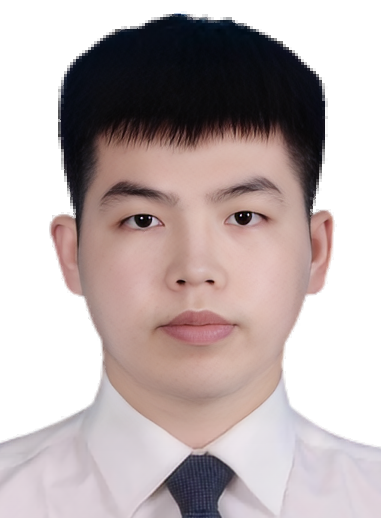}}]
{Zhaokang~Yin} received a master's degree in Software Engineering from Sun Yat-sen University. His research interests mainly include Blockchain.
\end{IEEEbiography}

\begin{IEEEbiography}
[{\includegraphics[width=1in,height=1.25in, clip,keepaspectratio]{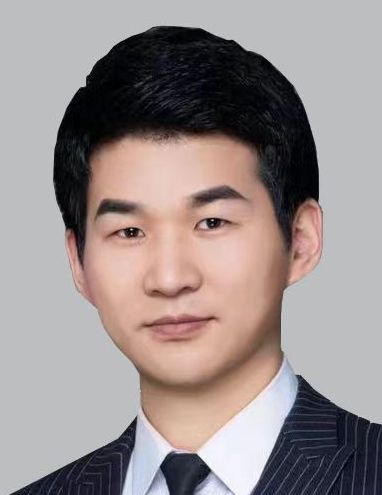}}]{Xiaofei~Luo} received a Ph.D. degree in computer science and engineering from the University of Aizu in March 2023. He is currently a postdoctoral researcher with the School of Software Engineering, Sun Yat-sen University, China. His research interests mainly include blockchain, payment channel networks, and reinforcement learning. His research has been published in the IEEE JSAC, TC, TPDS, and other well-known international journals and conferences.
\end{IEEEbiography}

\begin{IEEEbiography}
[{\includegraphics[width=1in,height=1.25in,clip,keepaspectratio]{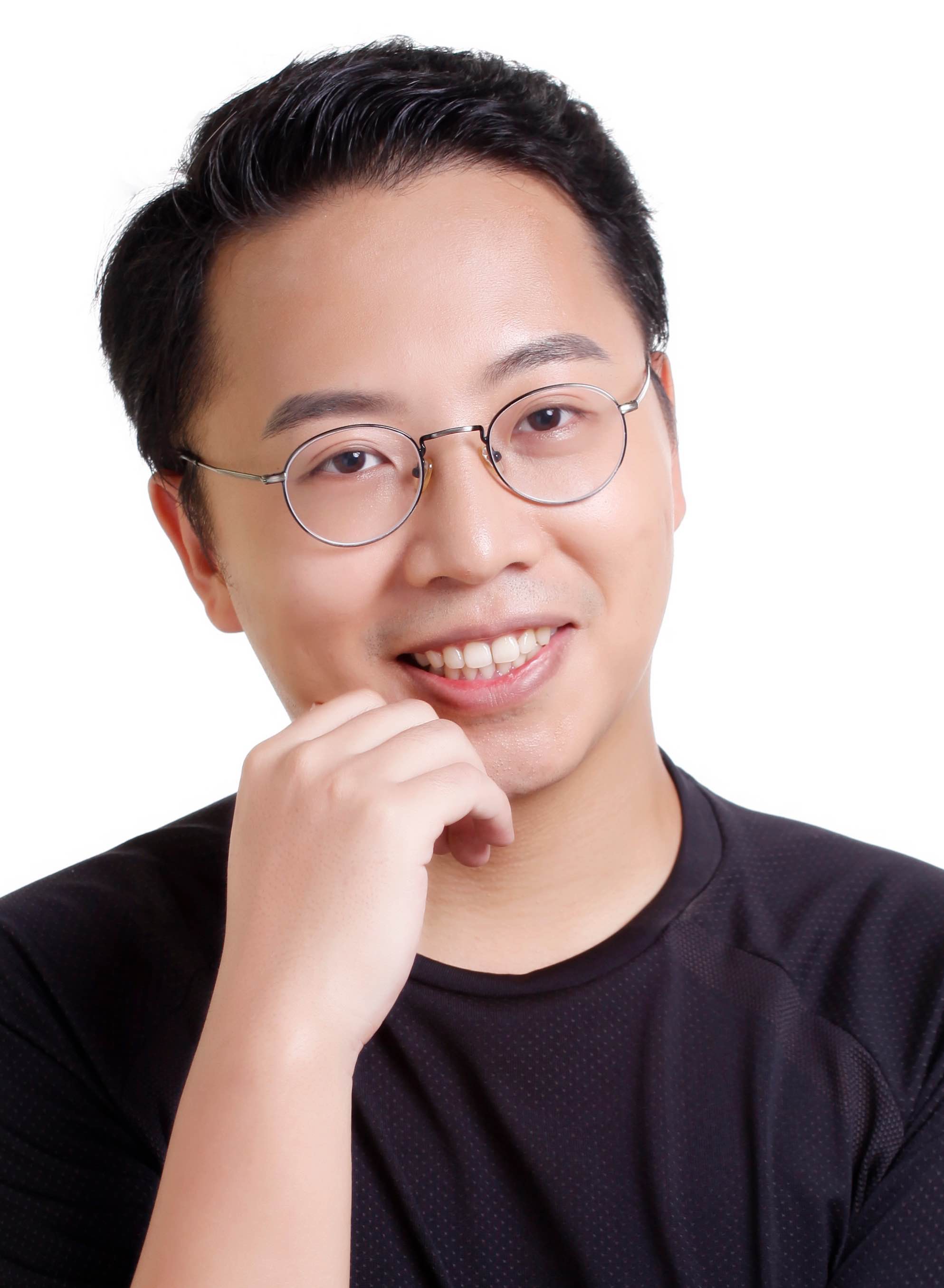}}]{Jianru~Lin} is a research scientist and engineer with HuangLab (a blockchain laboratory at Sun Yat-sen University). He has extensive experience designing and implementing decentralized systems, smart contract languages, and virtual machines. He translated the Chinese edition of Designing for Scalability with Erlang/OTP.
\end{IEEEbiography}

\begin{IEEEbiography}[{\includegraphics[width=1in,height=1.25in,clip,keepaspectratio]{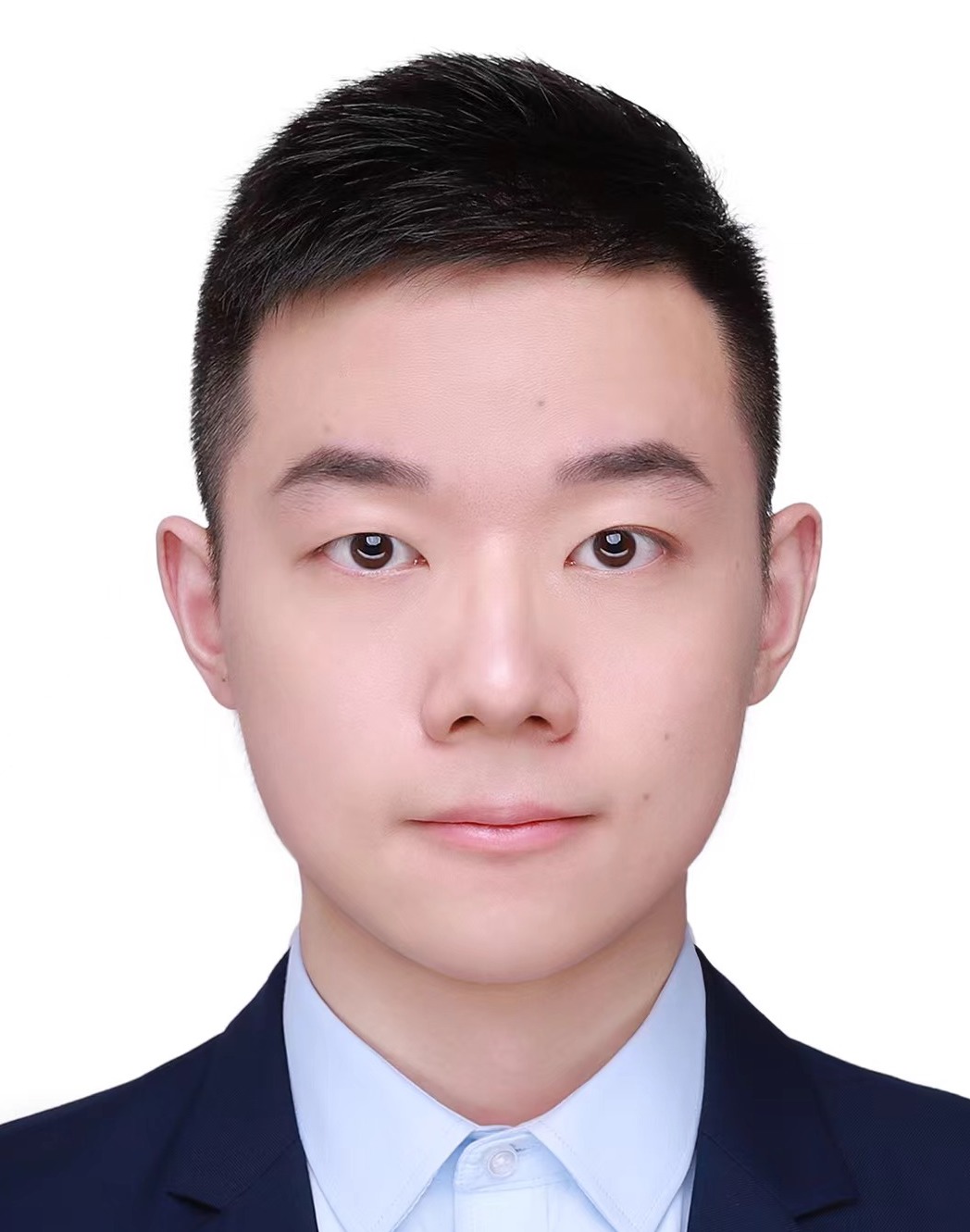}}]{Jian~Zheng} is currently working toward the PhD degree at Sun Yat-sen University, China. His research interests mainly include blockchain.
\end{IEEEbiography}

\begin{IEEEbiography}
[{\includegraphics[width=1in,height=1.25in,clip,keepaspectratio]{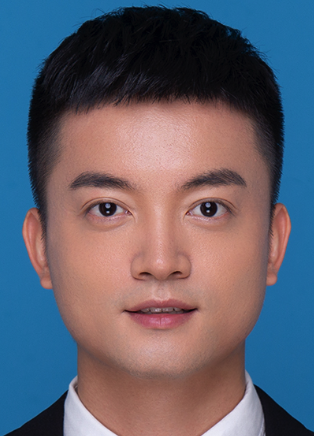}}]
{Taotao~Li} received the PhD degree in cybersecurity from the Institute of Information Engineering, Chinese Academy of Sciences and University of Chinese Academy of Sciences, China, in 2022. He is currently a postdoc with the School of Software Engineering, Sun Yat-sen University, Zhuhai, China. His main research interests include blockchain, Web3, and applied cryptography.
\end{IEEEbiography}

% \vfill
% \vspace{-2cm}

\begin{IEEEbiography}[{\includegraphics[width=1in,height=1.25in,clip,keepaspectratio]{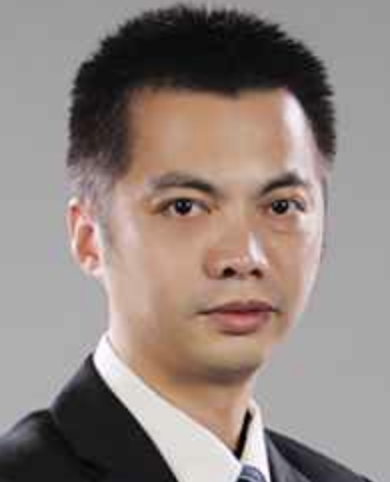}}]
{Zibin~Zheng} (Fellow, IEEE) received the PhD degree from the Chinese University of Hong Kong, Hong Kong, in 2012. He is a professor in the School of Software Engineering, Sun Yat-sen University, China. His current research interests include service computing, blockchain, and cloud computing. He received the Outstanding PhD Dissertation Award of the Chinese University of Hong Kong in 2012, the ACM SIGSOFT Distinguished Paper Award at ICSE in 2010, the Best Student Paper Award at ICWS2010, and the IBM PhD Fellowship Award in 2010. He served as a PC member for IEEE CLOUD, ICWS, SCC, ICSOC, and SOSE.
\end{IEEEbiography}

\end{document}